\documentclass[10pt,journal,compsoc]{IEEEtran}
\ifCLASSOPTIONcompsoc
  \usepackage[nocompress]{cite}
\else
  \usepackage{cite}
\fi
\ifCLASSINFOpdf
\else
\fi
\usepackage{url}
\usepackage{graphicx}	% For the graphics
\usepackage{caption}
\usepackage{subcaption}

\usepackage{pdflscape} 
\usepackage{longtable} 
\usepackage{colortbl}

\usepackage[utf8]{inputenc}
\usepackage{enumitem}
\usepackage{comment}

\newcommand{\mypar}[1]{\noindent{\textbf{#1:}}}

\newcommand{\question}[1]{\textit{#1}}

\newcommand{\new}[1]{\textcolor{blue}{#1}}
\renewcommand{\new}[1]{#1}

\newcommand{\tablefirstrowformat}[1]{{\textbf{#1}}}%\scriptsize %\small
\newcommand{\tabletextsize}[1]{{#1}} %\small %\scriptsize

\newcommand{\secstyle}[0]{Section} 
\newcommand{\figstyle}[0]{Figure} 
\newcommand{\tabstyle}[0]{Table} 
\newcommand{\kavagaitshort}[0]{KAVAGait} 
\newcommand{\kavagait}[0]{KAVAGait } 
\newcommand{\suppmat}[0]{(\url{http://phaidra.fhstp.ac.at/o:1928})}
\newcommand{\mli}[1]{\mathit{#1}}

\usepackage{todonotes}  
\newcommand{\sout}[1]{}

\begin{document}

%\title{KAVAGait: A Knowledge-Assisted Visual Analytics System for Clinical Gait Analysis}
\title{%
%KAVAGait -- Knowledge-Assisted Visual Analytics Solution for Gait Analysis: A Design Study\\
%Knowledge-Assisted Visual Analytics for Gait Analysis: The KAVAGait Design Study\\
%KAVAGait: A Design Study in Knowledge-Assisted Visual Analytics for Gait Analysis\\
%Knowledge-Assisted Visual Analytics\\for Gait Analysis: A Design Study
KAVAGait: Knowledge-Assisted Visual Analytics\\for Clinical Gait Analysis
}


%\author{Michael~Shell,~\IEEEmembership{Member,~IEEE,}
%        John~Doe,~\IEEEmembership{Fellow,~OSA,}
%        and~Jane~Doe,~\IEEEmembership{Life~Fellow,~IEEE}% <-this % stops a space
\author{Markus Wagner, Djordje Slijepcevic, Brian Horsak,\\ Alexander Rind, Matthias Zeppelzauer, and Wolfgang Aigner% <-this % stops a space
\IEEEcompsocitemizethanks{\IEEEcompsocthanksitem Markus Wagner, Djordje Slijepcevic, Alexander Rind, Matthias Zeppelzauer, and Wolfgang Aigner are with St. P\"olten University of Applied Sciences, Austria and TU Wien, Austria. E-mail: \{markus.wagner, djordje.slijepcevic, alexander.rind, matthias.zeppelzauer, wolfgang.aigner\}@fhstp.ac.at.
\IEEEcompsocthanksitem Brian Horsak is with St. P\"olten University of Applied Sciences, Austria. E-mail: brian.horsak@fhstp.ac.at.\protect\\
}% <-this % stops a space
\thanks{Manuscript received February 03, 2017; revised December 10, 2017.}}


\markboth{To Appear in IEEE TVCG}%
{Wagner \MakeLowercase{\textit{et al.}}: KAVAGait: Knowledge-Assisted Visual Analytics for Clinical Gait Analysis}
\IEEEtitleabstractindextext{%
\begin{abstract}
In 2014, more than 10 million people in the US were affected by an ambulatory disability. Thus, gait rehabilitation is a crucial part of health care systems.
The quantification of human locomotion enables clinicians to describe and analyze a patient’s gait performance in detail and allows them to base clinical decisions on objective data.
These assessments generate a vast amount of complex data which need to be interpreted in a short time period.
We conducted a design study in cooperation with gait analysis experts to develop a novel \textbf{K}nowledge-\textbf{A}ssisted \textbf{V}isual \textbf{A}nalytics solution for clinical \textbf{Gait} analysis (\kavagaitshort).
\kavagait allows the clinician to store and inspect complex data derived during clinical gait analysis. The system incorporates innovative and interactive visual interface concepts, which were developed based on the needs of clinicians. Additionally, an explicit knowledge store (EKS) allows externalization and storage of implicit knowledge from clinicians. It makes this information available for others, supporting the process of data inspection and clinical decision making.
We validated our system by conducting expert reviews, a user study, and a case study. Results suggest that \kavagait is able to support a clinician during %every day 
clinical practice by visualizing complex gait data and providing knowledge of other clinicians.
\end{abstract}

\begin{IEEEkeywords}
Design study, interface design, knowledge generation, knowledge-assisted, visualization, visual analytics, gait analysis.
\end{IEEEkeywords}}

\maketitle

% is not selected <OR> if conference mode is selected - because compsoc
% conference papers position the abstract like regular (non-compsoc)
\IEEEdisplaynontitleabstractindextext

% \fi
\IEEEpeerreviewmaketitle

%%%%%%%%%%%%%%%%%%%%%%%%%%%%%%%%%%%%%%%%%%%%%%%%%%%%%%%
%    Starting structural reorganizing by Markus W.    %
%%%%%%%%%%%%%%%%%%%%%%%%%%%%%%%%%%%%%%%%%%%%%%%%%%%%%%%

\ifCLASSOPTIONcompsoc
\IEEEraisesectionheading{\section{Introduction}\label{sec:introduction}}
\else
\section{Introduction}
\label{sec:introduction}
\fi
% for IEEE Computer Society journal papers produced under \LaTeX\ using
% IEEEtran.cls version 1.8b and later.
% % You must have at least 2 lines in the paragraph with the drop letter
% % (should never be an issue)
% I wish you the best of success.

% This is a biblio-test~\cite{munzner_nested_2009, sedlmair_design_2012}

\IEEEPARstart{A}{ccording} to the 2014 United States Disability Status Report ~\cite{erickson_2014_2016}, 5.5\% of working age adults (ages 21 to 64, amounting to more than 10 million nationwide) suffer from an ambulatory disability. %In addition, among the six types of disabilities identified, ambulatory  disability showed the highest prevalence rate of all. 
 Walking and stair-climbing are essential motor functions that are prerequisites for participation in activities of daily living. Disruptions to these motor skills hold severe health and socio-economic implications if left unattended. Therefore, gait rehabilitation is a crucial issue for clinicians. 

Gait analysis tools allow clinicians %(referred to as \textit{clinician} in this article) 
to describe and analyze a patient's gait performance to make objective, data based decisions. 
The systems commonly used for capturing gait data range from simple video cameras and force-distribution sensing walkways to highly sophisticated motion capture systems~\cite{nigg_biomechanics_2007,winter_biomechanics_2005}. The latter is often referred to as the gold standard in clinical gait analysis, as this method assesses the gait pattern's underlying kinematic and kinetic components~\cite{cappozzo_human_2005}. 
%However, 
However, the motion capture system's widespread use is limited due to its substantial monetary and infrastructural costs, prolonged time commitment for data collection, and its requirement for specialized technicians.
Thus, clinics with a large daily influx of patients must rely on more practical and affordable methods. Force plates and cost-effective two-dimensional gait analysis tools are popular alternatives to determine external forces applied to the ground (ground reaction force, GRF) during gait~\cite[pp. 83--96]{kirtley_clinical_2006} and the associated kinematic variables (e.g., 2D joint angles).
%Even though necessary and highly important for clinical observations,
%In detail, the individual load (vertical force component) and the shear forces (anterior-posterior and medio-lateral components) of each step can be measured~\cite[pp. 324--333]{nigg_biomechanics_2007}. The GRFs of a single step of a healthy individual resembles very specific waveforms. For clinical and research purpose several discrete parameters (e.g. local maxima and minima, loading rates, etc.) are derived from these waveforms~\cite{benedetti_data_1998}. In addition, by assessing a consecutive number of steps, spatio-temporal parameters (STP, e.g. step time and step length) are assessed~\cite[pp. 15--38]{kirtley_clinical_2006}. Both aspects are of great interest for the clinician, as gait impairments will have a strong impact on STP and on GRF. 
A typical clinical gait analysis scenario involves the following: in a first step, a clinician conducts a physical examination of the patient. Then, the patient is instructed to walk across a walkway in the gait laboratory several times, while the clinician records the patient's GRF. %gait data. 
%Resulting data are then typically composed to non-interactive visual representations such as line plots and simple spreadsheets to inform clinical decision making.
These analysis methods generate a vast amount of multivariate, time-oriented data, which need to be interpreted by the clinician in a short period of time.
However, support for decision making based on analyzing this data is very limited in currently used systems. The resulting data are typically represented in a very simplistic manner using non-interactive visual representations such as line plots and simple spreadsheets to inform clinical decision making.
%Let us think about John, who is a clinical gait analyst in a laboratory. If a patient comes to her, he/she firstly has to walk several times over a defined walkway, containing ground reaction force plates for measurements. Additionally, during the walks, the patient is filmed sideways to additionally support the process of clinical decision making. In the next step, the clinician uses the laboratory own system for gait data analysis. The system presents the collected data as waveforms (line plots) as well as several calculated additional parameters like spatio-temporal parameters (e.g., walk speed, step length, stride length). These data are visually represented in a non-interactive interface, whereby the waveforms are compared to a norm dataset and the parameters are represented as plain numbers. Based on these visual representations in combination with the clinician's work experience, clinical decision making is performed.

%Clinical decision making is a challenging process
%and automated data analysis methods can potentially support this 
%if clinicians' expertise can be integrated.
% Automated data analysis methods bear the potential to support the clinician during this challenging process.
In the above-described scenario, it is a difficult task to interpret the obtained data as several parameters are inter-linked and data interpretation requires considerable domain expertise.
The combination of a vast amount of inter-linked clinical data derived from clinical examinations, the need for sophisticated data analysis methods, and clinical decision making requiring the judgment and expertise of clinicians, strongly lends itself to the notion of visual analytics (VA)~\cite{thomas_illuminating_2005,keim_mastering_2010}.   
% * <tsiragy@gmail.com> 2017-02-02T00:06:54.278Z:
% 
% The paragraph above ("Automated data analysis...")  needs more of an explanation. I recommend defining what you mean exactly "automated data analysis methods" are.  Although seemingly obvious, it also sounds like this is a specific term that is used and therefore should be explained a bit further. 
% 
%Because the fact that manual analysis by domain experts is very cumbersome, automated data analysis methods are needed. In order to automate this process as much as possible, time-distance parameter (e.g.,~\cite{tahir2012parkinson}) ranges for particular `gait interferences' in comparison to the `normal gait' are needed  to be specified and categorized. This way, time-distance parameter from a patient can then semi-automatically be analyzed and matched in this context.
%On the other hand, this process cannot be automated completely as domain experts need to be in the loop to identify, correct, and disambiguate intermediate results. This combination of large amounts of data, complex data analysis needs, and the combination of automated data analysis with analytical reasoning by domain experts lends itself very well to the notion of visual analytics (VA) \cite{thomas_illuminating_2005,keim_mastering_2010}.
VA may support the clinician with powerful interactive visualization techniques that are integrated in combination with semi-automated data analysis methods. 
% * <tsiragy@gmail.com> 2017-02-02T00:08:44.742Z:
% 
% Same for "VA/visual analytics", I highly recommend defining what that is exactly to give more clarity to the reader. 
% 
% ^.
% for the purpose of exploratory data analyses and data interpretation.
% AR manual -> self-directed, exploratory, ...? or maybe leave it away
Consequently, this may support the clinician in interpreting complex data and drawing appropriate clinical conclusions.
% `implicit knowledge'~\cite{chen_data_2009}, sometimes also referred to as `tacit knowledge'~\cite{wang_defining_2009}, which in this case is the prior obtained experience by the clinician. 
The clinicians' `implicit knowledge' from prior experience is essential in the analysis process. %but
%So far it has not been shared with other experts or integrated in a VA system.
Thus, it %is logical
makes sense to externalize some of the domain experts' `implicit knowledge' and make it available as `explicit knowledge' in the VA process \cite{chen_data_2009, wang_defining_2009}.
%it is logical to externalize parts of this valuable information to make it available as `explicit knowledge' in the VA process \cite{chen_data_2009, wang_defining_2009}. 
% * <tsiragy@gmail.com> 2017-02-02T00:09:42.198Z:
% 
% Are "implicit and explicit knowledge" actual terms in the literature?  If yes, then they also need to be defined in more detail. 
% 
% ^.
%In VA the user (e.g. domain expert) is put into the loop and powerful interactive visualization techniques are employed to support manual data analysis. %In addition to challenging analysis methods, `implicit knowledge'~\cite{chen_data_2009} or `tacit knowledge'~\cite{wang_defining_2009} about the data, the domain experience or prior experience are often required to make sense of the data and not become overwhelmed.
% By externalizing some of the domain experts' implicit knowledge, it can be made available as explicit knowledge. 
As such, it can be used	 to augment the visual display of data and to support (semi-)automated data analysis (knowledge-assisted VA methods). 
Additionally, joint learning between clinicians %is 
would be enabled %and 
as well as the collection of expert knowledge across several clinicians allows the constructing of a comprehensive clinical knowledge database~\cite{chen_data_2009} (refereed to as `explicit knowledge store' (EKS) for the proposed prototype later in this article).

\subsection{Overview and Method}

This work follows the paradigm of problem-oriented research~\cite{sedlmair_design_2012}, i.e., working with real users (clinicians), and aims at solving the aforementioned problem %which originates in clinical practice 
by means of VA. 
%In detail, the developed methods should support the clinician during everyday clinical practice in interpreting gait data. 
In detail, a comprehensive prototype was developed which is intended to support the clinician in interpreting gait data during everyday clinical practice. 
The methods proposed in this work %should also be capable of 
aim at externalizing implicit knowledge of clinicians into a knowledge database that makes these data available as explicit knowledge to other clinicians. % and users. 
For this purpose, we conducted a design study, following Sedlmair et al.~\cite{sedlmair_design_2012}. Specifically, we followed the `nested model for visualization design and validation' %as proposed by Munzner
\cite{munzner_nested_2009}. This model is a unified approach, which structures visualization design into four levels and combines them with appropriate validation methods, which reduces threats to validity at each level.
%
%To be able to support domain experts in their work on clinical gait analysis, it is imperative to perform a design study~\cite{sedlmair_design_2012}. In particular, we follow the `nested model for visualization design and validation' as proposed by Munzner \cite{munzner_nested_2009} which is a unified approach structuring visualization design into four levels combining them with appropriate validation methods to reduce threats to validity at each level.
%Starting from the top, the levels are
%`domain problem and data characterization'%(understanding problem domain and users' tasks and goals)
%,
%`operation and data type abstraction'%(transform domain-specific description and raw data into a more generic notion)
%,
%`visual encoding and interaction design'%(design/matching of appropriate methods to represent and interact with data)
%, and `algorithm design'%(design of concrete implementation of methods).
%
Overall, our work contributes to visualization research in all three categories (1--3) outlined for design studies in \cite{sedlmair_design_2012} as well as presents new knowledge-assisted visualization approaches (4):
% by Sedlmair et al.~

\begin{enumerate}
\item \textbf{Problem characterization and abstraction:} A common language and understanding between domain experts and VA researchers was established.
Specific data, user, and task requirements for clinical gait analysis were set as prerequisites during the design process (see \secstyle~\ref{sec:backgroundandproblemcharacterization}) for the development of a novel knowledge-assisted VA system.
%% MW: Geändert auf Anraten von MZ.
%\item During the process of \textbf{problem characterization and abstraction}, common language and understanding between domain experts (clinicians) and visualization researchers was established.
%\secstyle~\ref{whatEver} summarizes the background of clinical gait analysis and condenses specifics of data, users, and tasks to be considered in VA design.
%The requirements gathered provide the basis, against which design proposals can be judged.

%The requirements gathered provide the basis, against which design proposals can be judged. %Thereby, the abstraction from the concrete domain vocabulary to the vocabulary of visualization also helps to match known visualization solutions from other domains to the problem at hand and vice versa serves as valuable basis for researchers other than those who have conducted the research.
\item \textbf{Validated design:} The design rationale and implementation details (see \secstyle~\ref{sec:designandimplementation}) were validated by conducting expert reviews, user studies, and a case study (see \secstyle~\ref{sec:validationandresults}). The primary aim was to obtain information whether or not the developed system is of valuable contribution to the domain.
%  identify the system's main benefits to the domain problem and to gain further knowledge about possible future development directions}. %were performed to validate the proposed prototype.
%of \kavagaitshort, presenting its design rationale and relevant implementation details (see \secstyle~\ref{whatEver}), expert reviews and user studies (see \secstyle~\ref{whatEver}), and a case study (see \secstyle~\ref{whatEver}).
%% MW: Umformuliert basierend auf den Anmerkungen von MZ.
%\item The \textbf{validated design} of \kavagaitshort{} demonstrates an effective VA solution for clinical gait analysis.
%\secstyle~\ref{whatEver} lays out its design rationale and relevant implementation details,
%\secstyle~\ref{whatEver} reports on expert reviews and user studies, and
%\secstyle~\ref{whatEver} presents a case study.

% Second, we `validate the visualization design' of the new designed and implemented `knowledge-assisted visual analytics gait analysis' prototype based on expert reviews, user studies and a case study. %This way, we test if the used visual data representations are effective and helpful to support the domain experts while solving their analysis tasks (domain problem).
\item \textbf{Reflection:} %analyzing our 
Insights gained during the validation process (see \secstyle~\ref{sec:reflectionandconclusion}) were reflected and analyzed
to propose directions for possible future development.
%\new{to gain further knowledge about possible future development directions.}
%for future improvements \new{to identify further needs for clinical gait analysis experts}. 
%Based on lessons learned and the transferability to other domains, other researchers might also propose different solutions based on this groundwork.
%% MW: Umformuliert basierend auf den Anmerkungen von MZ.
%\item We \textbf{reflect} in \secstyle~\ref{whatEver} on the design study ex post facto analyzing our insights gained during the performed user studies and case study. Based on this insights gained, we describe our lessons learned and the transferability to other domains. Thus, other researchers might also propose different solutions based on this groundwork.
\item New \textbf{knowledge-assisted visualization approaches} were used to generate easily understandable `Graphical Summaries' of the data as well as the novel `Interactive Twin Box Plots' (ITBP). %providing parameter based intercategory comparisons.  
\end{enumerate}

%As shown in the survey by Lam et al. \cite{lam_empirical_2012}, such contributions are still very rare. 
%Due to that and the importance of these steps as well as the fact that these stages are difficult and time consuming to carry out properly, researchers in the visualization community encourage to accept more papers on design studies and argue for considering such work to be first-class paper contributions in visualization venues \cite{munzner_nested_2009,sedlmair_design_2012}. 
%We aim for providing such a contribution in the domain of clinical gait analysis practice.

%In general, the stages included in a design study are difficult and time consuming to carry out properly.
%Such results are highly relevant contributions for the visualization community~\cite{munzner_nested_2009,sedlmair_design_2012}.
%Additionally, the gained knowledge can also be used and adapted to solve problem settings of related or other domains. 
%With this paper, we aim at providing such contributions for the domain of clinical gait analysis practice.

%%%%%%%%%%%%%%%%%%%%%%%%%%%%%%%%%%%%%%%%%%%%%%%%%%%%%%%

\section{Related Work}
\label{sec:relatedwork}
From a data perspective, gait measurements are multivariate time series. To visualize and analyze such data, a variety of different visual analytics (VA) approaches have been introduced in earlier work.

\mypar{Visual Analytics for Movement Time Series} Andrienko et al.~\cite{andrienko_visual_2013} give a broad overview how VA can be used to visualize locomotion, which they refer to as `Visual Analytics of Movement'. In their work, they give recommendations on how such data can be represented in the context of VA and how these data may be resampled. However, they mostly focus on geospatial datasets in relation to time.
In the field of sport science, three VA systems \cite{perin_2013_soccerstories, janetzko_feature-driven_2014, stein_2018_bring} support
soccer analysts in analyzing position-based soccer data at various levels of detail.
Janetzko et al.~\cite{janetzko_feature-driven_2014} additionally enrich the analysis with manually annotated  events such as fouls and suggest further candidate events based on classification.
Chung et al.~\cite{chung_2016_knowledge-assisted} applied a knowledge-assisted visual analytics approach to rank events in rugby plays.
An effective fully automated method for human motion sequence segmentation for character animation purposes was introduced by V\"ogele et al.~\cite{vogele_efficient_2014}. They described the fast detection of repetitions in discovered activity segments as a decisive problem of motion processing pipelines. For testing this method, they used different motion capture databases and visualized the results with stacked bar charts for comparison with other techniques.
In the context of medicine, sports, and animation, the `MotionExplorer' system~\cite{bernard_motionexplorer:_2013} enables the exploration of large motion capture data collections represented as multivariate time series.
Following an iterative design approach, Bernard et al.~\cite{bernard_motionexplorer:_2013} demonstrated the functionality of the `MotionExplorer' through case studies with five domain experts. 
A similar approach, the `MotionFlow' system~\cite{jang2016}, allows more specific grouping and  analysis of patterns in motion sequences.
Another system was described by Purwantiningsih et al.~\cite{purwantiningsih_visual_2016}. They collected data on patients' quality of movement using serious games and different motion sensing devices.
To make these multivariate time-series data accessible to clinicians, their VA solution allows hierarchical clustering and navigation in time.
The VA system `FuryExplorer'~\cite{wilhelm_furyexplorer:_2015} improves analytical workflows for evaluation of horse motion by interactive exploration of captured multivariate time-oriented data.

\mypar{Visual Analytics for Multivariate Time Series}
The analysis of time-oriented data is an important problem for many other domains beyond movement data.
In a systematic review, Aigner et al.~\cite{aigner_visualization_2011} surveyed more than 100 visualization approaches and systems for time-oriented data.
Many approaches for visualizing multivariate time series are based on a form of small multiples \cite{tufte_1983_visual} where the many charts -- one for each univariate time series -- are juxtaposed on a common time axis.
Space-efficient visualization techniques like horizon graph~\cite{heer_sizing_2009}, braided graph~\cite{javed_2010_graphical}, and qualizon graph~\cite{federico_2014_qualizon} have been designed and experimentally evaluated for such purposes.
The `LiveRAC' sytem~\cite{mclachlan_liverac:_2008} visualizes time series for hundreds of parameters in a reorderable matrix of charts, for IT systems management. The system allows for the reordering and side-by-side comparison with different levels of detail.
`KAMAS'~\cite{wagner_kamas_2017} is a knowledge-assisted visual malware analysis system, supporting IT-security experts during behavior-based malware analysis based on multivariate log files of the executed system and API calls of an operating system. %To evaluate the system, the authors performed focus group meetings, expert reviews and user studies.

The `PieceStack' system~\cite{wu_2016_piecestack} provides an interactive environment to split and hierarchically aggregate time series based on stacked graphs~\cite{havre_2002_themeriver}.
% Gnaeus~\cite{federico_gnaeus_2015} provides knowledge-assisted visualizations based on guidelines for electronic health records of multivariate time series data.
`Gnaeus'~\cite{federico_gnaeus_2015} provides visualizations of multivariate time series data from electronic health records using clinical guidelines for knowledge-assisted aggregation and abstraction.
A different approach to tackle multivariate data applies dimensionality reduction to project multivariate measurements to 2D space, where they can be displayed as trajectories (such as a connected scatter plot) \cite{haroz_2016_connected}.
%
%In many application domain, the analysis of time-oriented data is an important problem. There are many existing approaches for univariate time series and much viewer for multivariate time series to find pattern of interest in the data and help to understand them~\cite{bernard_timeseriespaths_2012}.
The `TimeSeriesPaths' system~\cite{bernard_timeseriespaths_2012} applied this approach as a visual data aggregation metaphor to group similar data elements.
Based on their VA approach, they provided a hatching based bar visualization for inner class comparison.
Schreck et al.~\cite{schreck_visual_2009} showed trajectories in small multiples and applied self-organizing maps to spatially cluster the trajectories.
%
% AR: Time Curves \cite{bach} might not related so closely, since it is based on similarity instead of multivariate data???
%
% possible further work: Tominiski time wheel;  Ordonez MTSA stardinates, Boegl TiMoVA
%
%In contrast, \kavagait is designed to support `interclass' comparison based on multivariate time series.
%
% Additionally, Heer et al.~\cite{heer_sizing_2009} compared the visualization of time series data in relation to line graphs and horizon graphs.
%
Other recent work relating to visual analytics of biomedical models include
PROACT~\cite{hakone_2017_proact}, PhenoLines~\cite{glueck_2018_phenolines}, and
a study on trust-augmenting designs~\cite{dasgupta_2017_familiarity}.
%

%\mypar{Summary}
The presented work focuses on multivariate time series data to solve problems in different domains. 
However, only one of the identified approaches (KAMAS~\cite{wagner_kamas_2017}) provide the ability to extract and store implicit knowledge of experts in the form of explicit knowledge in a database. 
This is a desirable feature for a VA tool, especially in clinical gait analysis, as it would support clinicians in decision making when analyzing a patient's gait and would support joint learning between different clinical experts.
%Gnaeus~\cite{federico_gnaeus_2015} uses guidelines for knowledge-assistance. Additionally, the soccer data explorer by Janetzko et al.~\cite{janetzko_feature-driven_2014} uses a `feedback loop' to tag unknown events which are used as training data (knowledge) for the systems automated classifier, but it does not support direct interactive knowledge exploration and comparison.  

%%%%%%%%%%%%%%%%%%%%%%%%%%%%%%%%%%%%%%%%%%%%%%%%%%%%%%%

\section{Problem Characterization \\ \indent ~\& Abstraction}
\label{sec:backgroundandproblemcharacterization}
One primary goal in clinical decision making during gait rehabilitation is to assess whether a recorded gait measurement displays normal gait behavior or if not, which specific gait patterns (abnormalities) are present.
To understand how to support the analysts in this context, we performed a `problem characterization and abstraction', defined as the first contribution of a design study~\cite{sedlmair_design_2012}. 
To ensure knowledgeable results for the domain of clinical gait analysis and rehabilitation, along the triangle of data, users and tasks \cite{miksch_matter_2014},
we followed a user-centered design process \cite{sharp_interaction_2007}.
Information was gathered primarily from focus group meetings~\cite[pp. 192]{lazar_research_2010} and set in context with domain-specific literature. 
Based on this, we addressed the first (\textit{domain problem and data characterization}) and second level (\textit{operation and data type abstraction}) of the nested model by Munzner~\cite{munzner_nested_2009}.

\subsection{Focus Group}
\label{subsec:focusgroup}
The primary aim of the focus group meetings was to match the domain-specific vocabulary between the computer scientists and clinical experts. 
In addition, these meetings were used to establish a mutual understanding of the following questions for the specific setting: 

\begin{itemize}
	\item What is the workflow in a clinical gait laboratory?
    \item How does the clinician interact within this setting?
\end{itemize}

\subsubsection{Focus Group Setup}
\mypar{Participants and Clinical Partner}  
Seven participants comprised the focus group (two clinical gait analysis experts, two pattern recognition experts, and three %knowledge-assisted 
visual analytics (VA) experts). \new{In addition the AUVA, as the mandatory social insurance for occupational risks for more than 3.3 million employees and 1.4 million pupils and students in Austria, served as a cooperation partner during the entire project. The AUVA runs several rehabilitation centers in Austria. The prototype described in this manuscript was developed along the needs of the AUVA's clinical gait laboratories and clinical practice.}

\mypar{Design \& Procedure} The focus group members shared a co-working space so that short stand-up meetings were possible and questions could be resolved quickly. Additionally, six focus group meetings with a duration of approximately one hour were held to discuss detailed questions\sout{ with all members}. All these activities were held over a 13-months time frame.

\mypar{Apparatus \& Materials} 
The results of the frequent discussions and meetings were regularly documented, which resulted in an extensive basis for a common mutual understanding. These notes were subsequently transformed into the manuscript at hand. 

\subsubsection{Focus Group Findings} %of the Focus Group}

%\begin{itemize}

\new{
The findings of the focus group sessions mainly concerned data-related aspects as well as a deeper understanding of the overall analysis process. Next, these findings are briefly summarized.}

\new{A sufficient amount of patient gait data is necessary to develop visualization and pattern recognition applications for the clinical practice. While there have been attempts to provide such gait analysis databases~\cite{tirosh_gaitabase:_2010}, the amount of publicly available data is still too limited.}
\new{The AUVA's rehabilitation centers typically use force plates to determine ground reaction forces (GRFs) to assess patient gait disorders and to evaluate patient progress during physical therapy treatment. 
The data used\sout{in this project} were acquired retrospectively from the AUVA's database.} %Thus, for the proposed developments only a subset of the available data were used.

\begin{figure}[t!]
  	\centering
  		\includegraphics[width=1\columnwidth]{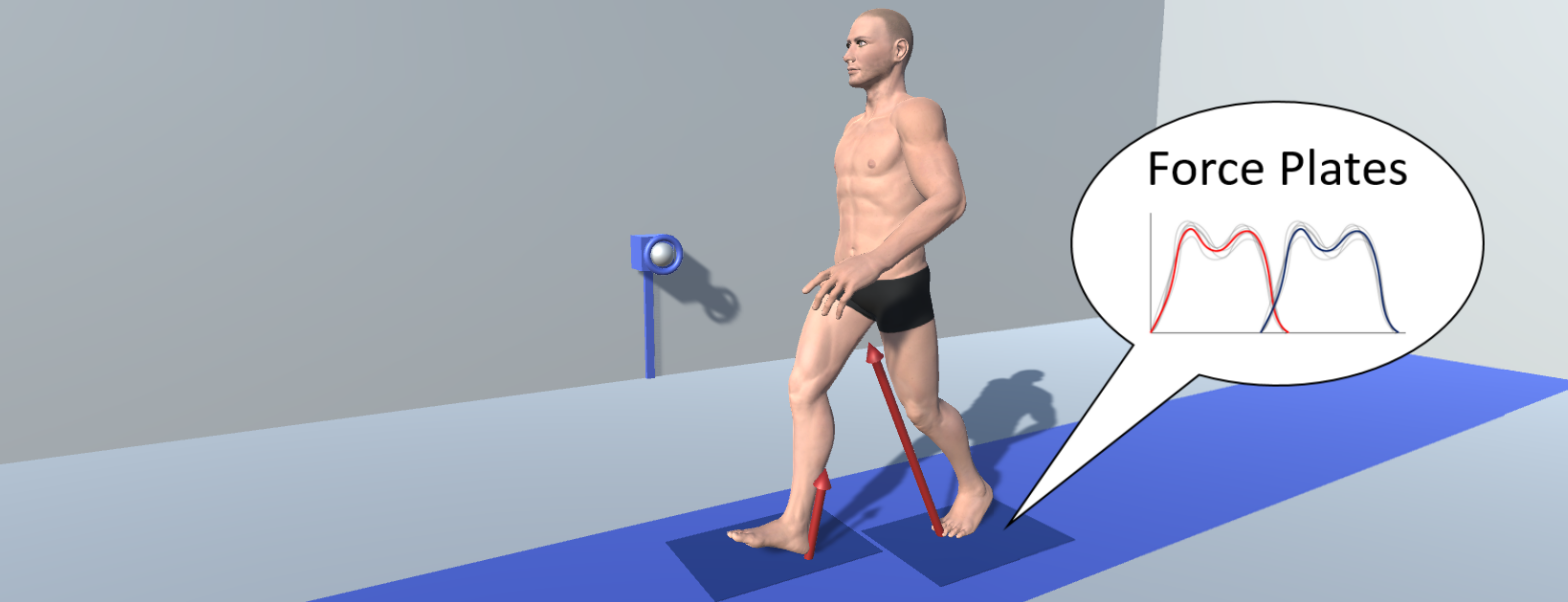}
  	\caption{During clinical gait analysis, two force plates integrated flush into the walkway allow for quantifying the ground reaction force (GRF) per foot during walking.}
	\label{fig:gaitanalysisprocedure}
\end{figure}

\begin{figure}[t!]
  	\centering
  		\includegraphics[width=1\columnwidth]{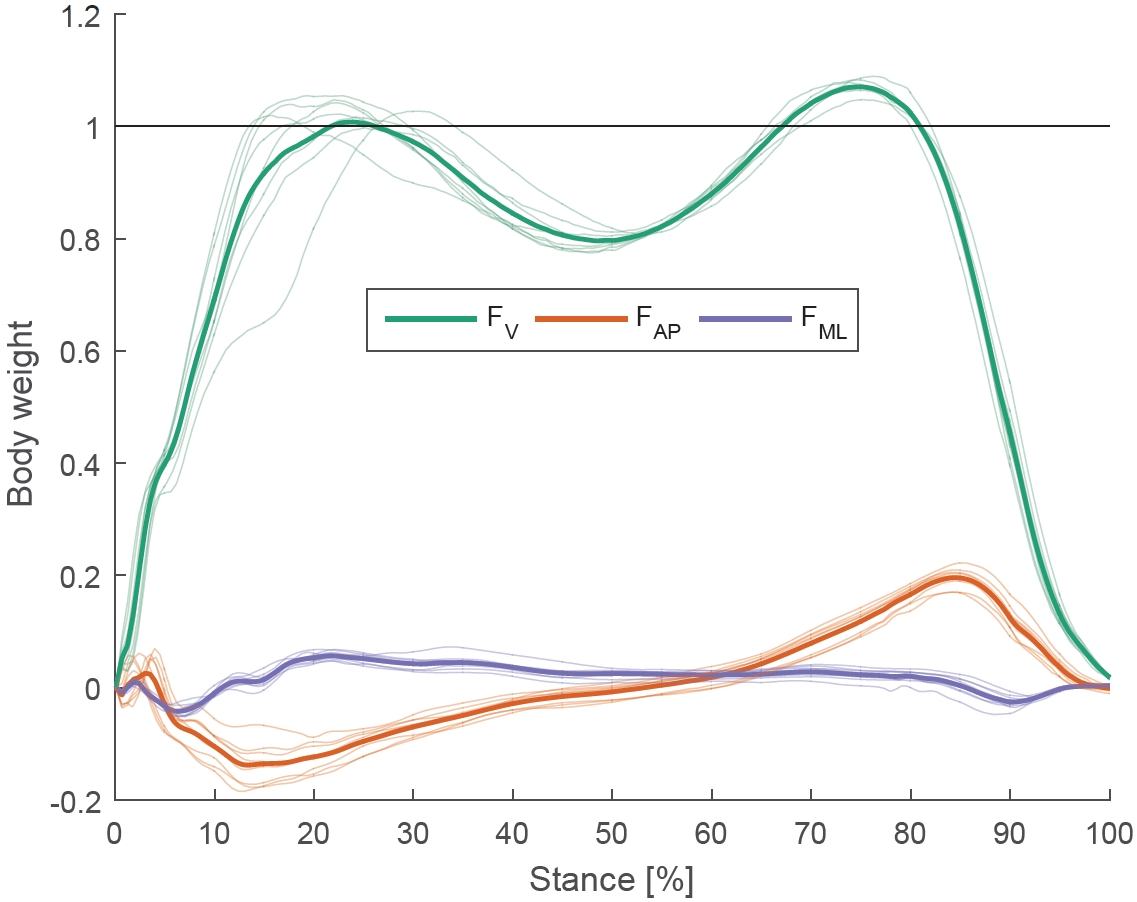}
  	\caption{Typical recordings (consistency graphs) of ground reaction force (GRF) data during a clinical gait analysis session. A total of 10 steps of the same foot are recorded, time-normalized to the stance phase of one step and amplitude-normalized to body mass. The three-dimensional GRFs are presented as the vertical force component ($F_V$) and the anterior-posterior ($F_\mli{AP}$) and medio-lateral ($F_\mli{ML}$) shear forces (\textit{the average curves are drawn in bold}). 
    }
	\label{fig:normalgaitgrfs} 
\end{figure}

\new{A typical gait analysis scenario may be divided into three main workflow stages:
% \begin{inparaenum} \item
(1) Rigorous physical examination of the patient by the clinician.
(2) Instruction of the patient during gait analysis and data recording:
the clinician guides the patient through the entire process of gait analysis
and instructs the patient to walk repeatedly across an approximately 10 meter walkway (see \figstyle~\ref{fig:gaitanalysisprocedure}).
In the center of the walkway, one or more force plates with an approximate size of 0.4\,x\,0.6 meters are fitted flush to the ground.
The clinician records necessary data by operating the measurement equipment and takes care that several clean footsteps and corresponding videos are recorded. 
(3) Processing and interpretation of the acquired data:
the clinician  processes recorded data using commercial software provided by the manufacturer of the measurement equipment.
%These systems typically present the collected data in a non-interactive interface, as simple waveforms (line plots) as well as several additionally calculated discrete parameters (e.g. walking velocity, step length, etc.) as plain numbers in tables and spread sheets (see \figstyle~\ref{fig:normalgaitgrfs}).}
These systems typically present the collected data in a non-interactive interface, as line plots of GRFs (see \figstyle~\ref{fig:normalgaitgrfs}) and several calculated discrete parameters (e.g., walking velocity, step length, etc.) as numbers in a table.}

\subsection{Data--Users--Tasks Analysis}
\label{subsec:datauserstasks}

%Above we have characterized the domain problem of clinical gait analysis
Above we have described the general  workflow of clinical gait analysis. %These information were derived from focus group meetings.
In addition to these results, we 
% followed the approach of Miksch \& Aigner~\cite{miksch_matter_2014} to 
structured the `domain problem and data characterization'
%(defining the first stage of the \textit{nested model} by~\cite{munzner_nested_2009})
along the Data-Users-Tasks triangle \cite{miksch_matter_2014}, which will be described in the following% section
. This high-level framework is structured around three questions:

\begin{itemize}\itemsep0em
	\item What kind of data are the users working with? (\textit{data})
	\item Who are the users of the VA solution(s)? (\textit{users})
	\item What are the (general) tasks of the users? (\textit{tasks})~\cite{miksch_matter_2014}
\end{itemize}

%Based on 
The answers to these questions support designers of VA methods %are supported
to find or design appropriate visual representations of the data combined with appropriate analysis and interaction methods that support the users. % in their work.

\mypar{Data}
\new{The aforementioned force plates measure three components of the applied force as a vertical, anterior-posterior and medio-lateral force component, sampled typically at 1000 Hz \cite[pp. 325--326]{nigg_biomechanics_2007}.}
Even though all three components are necessary to describe a patient's gait pattern, the vertical component receives most attention in research and clinical practice.
The reason is that the vertical component shows greater magnitudes than the other two components~\cite{hamill_biomechanical_2006}.
% During walking for example the vertical GRF reaches a magnitude of 1 to 1.2 of body weight (BW), whereas the anterior-posterior and medio-lateral direction only show magnitudes of 0.15 and 0.01 of BW, respectively~\cite{hamill_biomechanical_2006}.
\new{To foster comparability of the GRF data between different patients, each GRF signal is amplitude-normalized by the product of body weight ($BW$) and standard gravity ($g_0$) and time-normalized by 
expressing the signals as 100\% of stance time.} Resulting data for each foot are then visualized by plotting so called consistency graphs, where all trials are plotted into one graph to inspect variability across the steps recorded (see \figstyle~\ref{fig:normalgaitgrfs}). %Then, for visual inspection simplicity, a mean representative curve from all trials and corresponding standard deviation bands are calculated and plotted. From these data several discrete biomechanical parameters can be derived, such as amplitudes and time points of local peaks and valleys.
Although GRF is a very sensitive measure of gait pathology, its specificity is low since GRF comprises the motion and acceleration of whole body dynamics~\cite[pp. 95]{kirtley_clinical_2006}. Thus, additional measurements are necessary to describe the gait pattern of an individual in detail. In clinical gait analysis, the repeated movement of steps are referred to as a gait cycle, which starts with the initial contact of one leg with the ground, to the next ground contact of the same leg.
Within this concept, one can assess spatial and temporal parameters %, referred to as spatio-temporal parameters 
(STPs) of gait~\cite{baker_measuring_2013}. Spatial parameters comprise the length of a step or a stride (two consecutive steps). Temporal parameters comprise the time duration of for example a single step, a stride, or the swing phase. Additionally, the cadence (steps per minute), number of gait cycles per specified time, and walking speed are used to express the temporal aspect of gait.

\mypar{Users} Clinical gait analysis is performed by domain experts -- \emph{physicians, physical therapists, bio-medical engineers, or movement scientists}.
%These users have a strong background in gait and movement analysis, typically holding a university degree.
% They possess background knowledge about anatomy, biomechanics, gait analysis and a particular intuition on pathological gait functions.
They possess background knowledge about functional anatomy, biomechanics, and gait analysis.
The users are comfortable using different data representations (e.g., spreadsheets, box plots, line plots), mostly developed for a special hardware setting. %for their institution.
Thus, %depending on the used software solutions, 
they have no dedicated experience with VA solutions. %Generally, gait analysis is a specialist skill that requires experience, and interdisciplinary knowledge from medical and technological domains.

\mypar{Tasks} The primary task of a clinician in gait rehabilitation is to \textit{assess gait performance, to analyze and interpret the acquired data and to use this information for clinical decision making}.
Secondary tasks involve the identification of specific gait patterns (abnormalities) and the comparison of observed data to already existing patient data sets (e.g., in the clinic's database). To support these tasks, expert knowledge might be stored in some sort of database, so that this information can be shared with other clinicians.

\subsection{Prototype Requirements}
\label{subsec:prototyperequirements}

Based on the insights gained in \secstyle~\ref{subsec:datauserstasks}, we defined four key requirements (R) which have to be fulfilled by the \kavagait system:

\begin{description}[labelindent=0.0cm,leftmargin=0.5cm]
	\item[\textbf{R1 Data:}] \textit{Handling of complex data structures in clinical gait analysis.} 
    To ensure the effective exploration and analysis, time-dependent \new{three-}dimensional ground reaction forces (GRFs) and %globally calculated 
    spatio-temporal parameters (STPs) need to be modeled, stored, and visualized for inspection. In addition, for clinical decision making a visualization of the patients' raw data is essential.
	
    \item[\textbf{R2 Visual Representation:}] \textit{Visual representations appropriate for gait analysis experts.} 
    Clinicians use different %types of 
    diagrams (e.g., box %plots 
    and line plots) to conduct their analyses.  
    
    \item[\textbf{R3 Workflow:}] \textit{Workflow-specific interaction techniques.} 
	%In relation to the clinical gait analysis workflow, 
    It is important to provide familiar interaction techniques\sout{and metaphors} (e.g., drag and drop, sorting, filtering) to the clinicians\sout{, which support} supporting the identification of specific gait patterns and the comparison to already existing data sets of patients. % for clinical decision making.
    
    \item[\textbf{R4 Expert Knowledge:}] 
    \textit{Externalization of expert knowledge to reuse and share.} 
    When analysts solve real world problems, they have a vast amount of data at their disposal to be analyzed and interpreted. % for clinical decision making. 
% By externalizing and storing of the experts' implicit knowledge~\cite{chen_data_2009}, it can be made system-internally available as computerized knowledge to support further analysis or other analysts.
By storing the clinicians' implicit knowledge, it can be made internally available in the system and usable to support the analysis process.
\end{description}

\noindent
%%% AR: if has to be shorter I would drop the colorful sentence part with "basic pillars". 
%%% AR: the second sentence part is repetition from before the list
These four requirements form the basic pillars of KAVAGait\sout{, which} \new{ and} have to be fulfilled during the design and implementation\sout{ phase}.

%%%%%%%%%%%%%%%%%%%%%%%%%%%%%%%%%%%%%%%%%%%%%%%%%%%%%%%

\section{Design \& Implementation}
\label{sec:designandimplementation}
\begin{figure*}[ht!]
  	\centering
  		\includegraphics[width=1\textwidth]{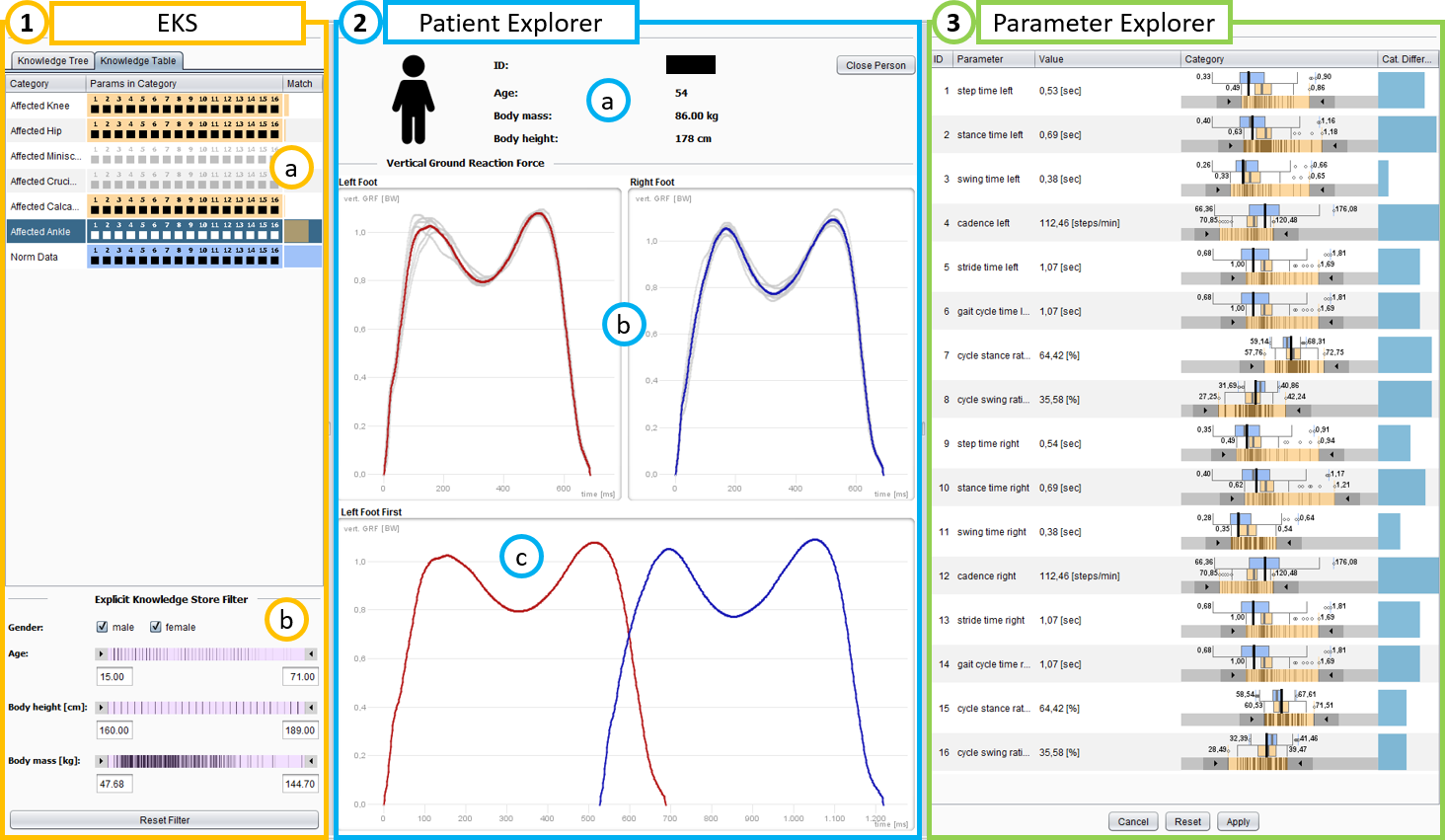}
  	\caption{User interface of \kavagait with its three main areas for gait analysis. 
    1) The table structure in (1a) shows the explicit knowledge store (EKS) that provides an overview of the stored gait patterns and how good the currently loaded data matches their definitions along with the controls in (1b) are used for filtering the included data in the EKS. 
    2) The patient explorer including the (2a) `Person Information', the (2b) visualization of the `vertical ground reaction force' (${F_{v}}$) for each individual foot and the (2c) visualization of the combined ${F_{v}}$ from both feet.
    3) Shows the `Parameter Explorer' visualizing the 16 calculated spatio-temporal parameters (STPs) of the loaded person in relation to the `Norm Data Category' and a second `Selected Category'.}
	\label{fig:prototypescreen1}
\end{figure*}

To keep the design in line with the needs and requirements defined earlier (see \secstyle~\ref{sec:backgroundandproblemcharacterization}), we continued our user-centered design process~\cite{sharp_interaction_2007} by involving three domain experts in clinical gait analysis.
% for design and implementation.
% To support clinical gait analysts during their work, we followed a user-centered design process~\cite{sharp_interaction_2007} for the algorithm design and implementation.
% Therefore, we involved a group of three domain experts in gait analysis to keep the design in line with the former analyzed needs (see \secstyle~\ref{sec:backgroundandproblemcharacterization}). % IN DER DISS MUISS AN DIESER STALLE AUF DIE REQUIREMENTS VERWIESEN WERDEN DIE DANN NOCH EXTRA ERARBEITET WERDEN AUF BASIS DER DATA_USERS_TASKS ANALYSE --> SIEHE KAMAS!
We iteratively produced sketches, screen prototypes, and functional prototypes~\cite{kulyk_human-centered_2007}.
Thus, we could gather and apply feedback about the design's usability and how well it supports  \new{clinicians'} needs\sout{of the clinicians}.
This way, we addressed the third (\textit{visual encoding and interaction design}) and fourth level (\textit{algorithm design}) of the nested model~\cite{munzner_nested_2009}.  
%
%
%\begin{figure*}[ht!]
%  	\centering
%  		\includegraphics[width=1\textwidth]{Figures/Prototype_Screen_2_OLD}
%  	\caption{User interface of the \kavagait system with its two areas (1 and 2) for EKS exploration and adjustment in relation to stored single `Patients', and (3 and 4) for EKS exploration and adjustment in relation to `Categories' containing several `Patients'.
%    1) The tree structure of the EKS while selecting a single `Patient' for comparison and adjustment 2) with other patients in relation to the `Norm Data Category' and the `Category' including the patient.
%    3) The tree structure of the EKS while selecting a `Category' for comparison 4) based on the hatching representing the time-distance parameters of the patients included in the category.
%    }
%	\label{fig:prototypescreen2}
%\end{figure*}
%
The design study resulted in \kavagaitshort{},\new{\sout{(see \figstyle{}s~\ref{fig:prototypescreen1},~\ref{fig:prototypescreen2}, and~\ref{fig:prototypescreen3})}} which is implemented in Java, based on a data-oriented design~\cite{fabian_data-oriented_2013} (e.g., used in game development and real-time rendering).
Next, we elaborate on central design decisions.

%\mypar{Internal Data Handling Concept}
%To increase the performance of our system, we decided to use a data-oriented design~\cite{fabian_data-oriented_2013} (e.g., used in game development and %real-time rendering) to organize and perform transformations on the loaded data.
%For example, we implemented map and set data structures as translation tables and for the input data, we mapped the strings to integers. 
%Based on this mapping, we built an internal data structure to increase performance and to decrease memory usage. 
%In combination with these structures, we implemented an action pipeline (similar to a render pipeline) in which it is possible to link filter options in order to realize high performance dynamic query environments and to perform all operations on the data in real-time. 

% \mypar{Data} The \kavagait system
\subsection{Input Data} The\sout{primary} input data for\sout{the} \kavagaitshort{}\sout{system} are the vertical component of \new{the} ground reaction force ($F_v$) of both feet collected by force plates \sout{in the form of}\new{as} two synchronized time series% combined in one import file
.
From these time series, spatio-temporal parameters (STPs) are calculated as 16 discrete numbers.
Additional patient data on gender, age, body mass, and body height are available.

\subsection{Explicit Knowledge Store (EKS)} 
To support gait analysts during their work, we designed the EKS related to STPs for different categories of gait patterns. %specific functional abnormalities and norm data of healthy gait.
Generally, the EKS stores the computerized form of the analysts' implicit knowledge as explicit knowledge. Therefore, categories of different gait patterns are used, whereby each contains 16 value ranges depending on the 16 STPs describing the gait analysis result of a patient.
For each category, the EKS contains the previously assigned patients and\sout{the} \new{their STP} values\sout{ of their STPs}.
Additionally, clinicians can refine the value range $[min, max]$ for each STP and category manually.

% Based on the EKS \dots 

%These parameters are 
% stored in the EKS as explicit knowledge to be used for automated analysis 
% represented as `graphical summary' and matching (see \figstyle~\ref{fig:prototypescreen1}:1a). %Additionally, the parameters are used for interclass comparisons provided by ITBPs (see \figstyle~\ref{fig:prototypescreen1}:3 and~\ref{fig:prototypescreen2}:2). %to support the analyst during clinical decision making.
% organized as categories of specific gait patterns (anomalies) and norm data of healthy gaits.

\subsection{Visual Interface Design Concept}
%In relation that physio therapists and gait analysis experts do not have to have programming skills, we tried to create a interface structure which allows to work from the left to the right to fulfill their analysis tasks.
 To best support physical therapists and gait analysis experts, we created an interface structure that allows working from \textit{left-to-right} to fulfill the analysis tasks.
This interface structure establishes an easy workflow concept based on multiple views for gait analysis experts. 
In relation to this interface structures, we situated the table structure of the EKS (see \figstyle~\ref{fig:prototypescreen1}:1) as well as the 
tree structure of the EKS (see \figstyle{}s~\ref{fig:prototypescreen2}:1 and~\ref{fig:prototypescreen3}:1) to the left side of the interface to select individuals or categories of interest for exploration that always includes the related filtering options.
%In general, \kavagait provides three different views, one for the exploration of new loaded patient data (see \figstyle~\ref{fig:prototypescreen1}) and two for the exploration and adjustment of the EKS (see \figstyle~\ref{fig:prototypescreen2}).  

\subsection{Visualization Concept}
\begin{figure*}[ht!]
  	\centering
  		\includegraphics[width=1\textwidth]{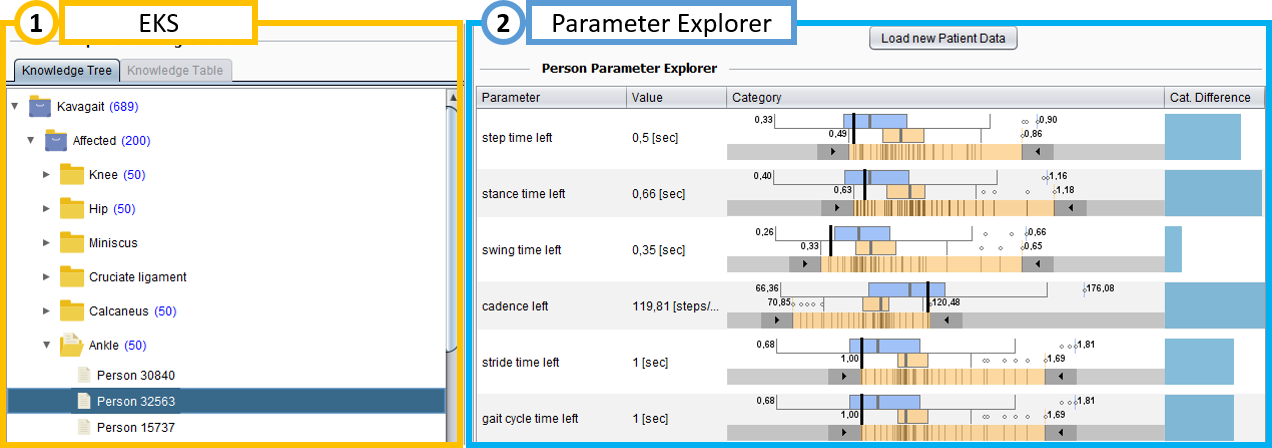}
  	\caption{User interface for `explicit knowledge store' (EKS) exploration and adjustment in relation to stored single patients.
1) The tree structure of the EKS while selecting a single patient for comparison and adjustment 2) with other patients in relation to the norm data category and the category that includes the patient (Ankle in this case).}
	\label{fig:prototypescreen2}
\end{figure*}

\begin{figure*}[ht!]
  	\centering
  		\includegraphics[width=1\textwidth]{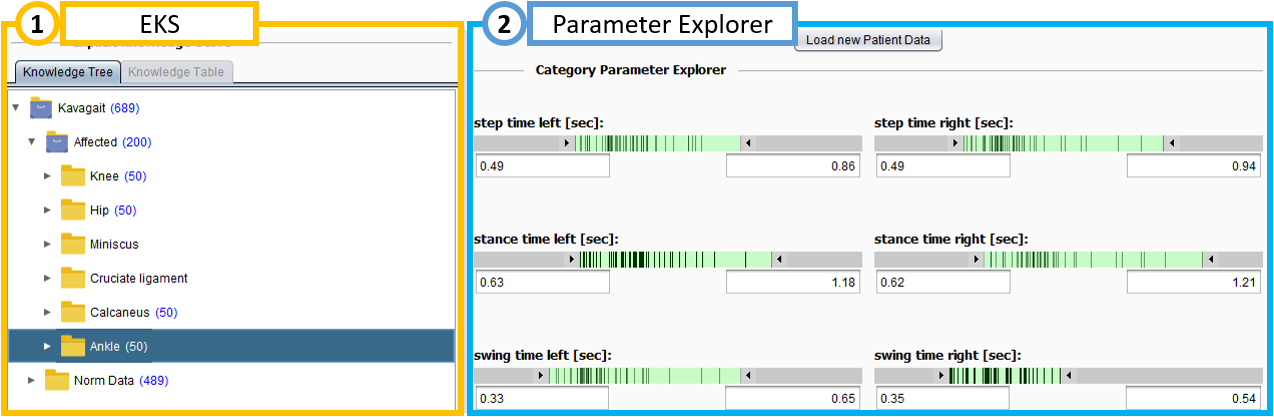}
  	\caption{User interface for `explicit knowledge store' (EKS) exploration and adjustment in relation to categories containing several patients.
1) The tree structure of the EKS while selecting a category (`Ankle' in this case) for \new{adjustment\sout{comparison}} 2) based on the 'Hatching Range Slider' (HRS) representing the limits or norm-ranges of the spatio-temporal parameters from the patients included in the category.}
	\label{fig:prototypescreen3}
\end{figure*}

The \kavagait system % provides %three different views to
supports two major use cases 1)~to  assess newly acquired patient gait data as elicitated in \secstyle~\ref{sec:backgroundandproblemcharacterization} \new{\sout{(see \figstyle~\ref{fig:prototypescreen1})}} and additionally 2) to explore and adjust the stored explicit knowledge \new{\sout{(see \figstyle{}s~\ref{fig:prototypescreen2} and~\ref{fig:prototypescreen3})}}.

The visual representations used in \kavagait have been developed through continuous refinement in multiple focus group sessions. Typically, we started from sketches based on a small number of known visualization techniques for multivariate time-oriented data~\cite{aigner_visualization_2011}. Those suggestions were then discussed in focus group meetings with the domain experts and continuously improved or new ones derived (e.g., ITBP) to fulfill the users’ needs.

\subsubsection{Assessment of Newly Loaded Patient Gait Data}
When loading new `Patient' data, the information \new{\sout{(see \figstyle~\ref{fig:prototypescreen1}:2a)}} containing the `ID', `Age', `Body mass', `Body height' and `Gender',
and the measurements of the `vertical ground reaction forces' ($F_v$) \new{\sout{(see \figstyle{}s~\ref{fig:prototypescreen1}:2b and 2c)}} are visualized in the center view of \kavagait (see \figstyle~\ref{fig:prototypescreen1}:2). These $F_v$ data are represented for each foot\new{\sout{(see \figstyle~\ref{fig:prototypescreen1}:2b)}}, whereby the light gray lines represent a single step and the red (left foot) or blue (right foot) line represent the mean $F_v$ data of the single steps. Additionally, a joint representation of $F_v$ on a combined temporal axis is available \new{\sout{(see \figstyle~\ref{fig:prototypescreen1}:2c)}} for further analysis and comparison.
To make the $F_v$ of a newly loaded patient comparable with others, they are normalized by body weight.

For identification of possibly matching gait patterns, % is possible in
the `Knowledge Table' (see \figstyle~\ref{fig:prototypescreen1}:1a) relates the newly loaded patient's calculated STPs to `Categories' (pathologies) of specific gait patterns (gait abnormalities) or norm data (describing healthy gait).
The\new{\sout{se}} 16 calculated STPs are the input for the visualizations % in the `Knowledge Table' (see \figstyle~\ref{fig:prototypescreen1}:1a)
in the `Params in Category' column.
Depending on the 16 STPs, the so called `Graphical Summary' tells the clinician if a patient parameter $x$ is in range $[min, max]$ with a black rectangle or if it is out of range {($x < min$ $\lor$ $x > max$)} with a black rectangular frame based on the calculated ranges out of the EKS. If the EKS does not contain data for a category (empty category), the `Graphical Summary' represents a gray rectangle.
Thus, these three states provide a first overview of the patient. 
The third column (`Match') represents how the newly loaded patient matches to the stored categories in the EKS (a wider bar means a better match) supporting the clinicians during clinical decision making.
For each category, a bar of width $c$ is computed according to Equation~\ref{math:matching}:

\begin{equation}
	c = \sum \limits_{i=1}^{n} \frac{\sigma_i}{max(|\mu_i - x_i|^2, \epsilon)}
\label{math:matching}
\end{equation}

\begin{figure}[t!]
	\centering
		\includegraphics[width=0.8\columnwidth]{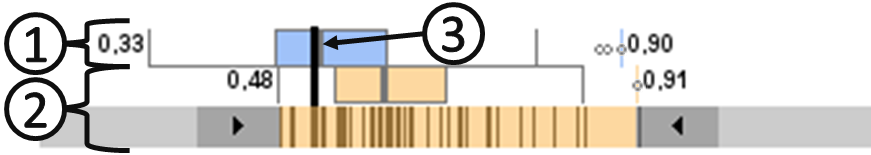}
		\caption{Illustration of the `Interactive Twin Box Plot' (ITBP) for intercategory comparison.
        1) represents the `Norm Data Category' as a blue box plot, 2) represents the `Selected Category' of a specific gait abnormality as an orange box plot in combination with a `Hatching Range Slider' (HRS) and 3) represents the actual STP values of the currently loaded patient for comparison.}
	\label{fig:interactivetwinboxplot}
\end{figure}

Equation~\ref{math:matching} defines the matching between a sample and a `Gait Category'.
%
%has to be used where $c$ defines the matching criteria results for the category, 
%
Thereby, $i$ iterates over all 16 STPs, $\sigma_i$ is the standard deviation and $\mu_i$ is the mean for the specific STP of all patients in the category. Additionally, $x_i$ is the specific STP of the newly loaded patient.
Note that Equation~\ref{math:matching} is an inverted variant of the Fisher criterion used in linear discriminant analysis (LDA) \cite{fisher1936use}. Variable $c$ grows with increasing agreement between $x_i$ and the distributions defined by $\mu_i$ and $\sigma_i$. $\epsilon$ is a small number that avoids potential division by zero.
By using the included filtering options (see \figstyle~\ref{fig:prototypescreen1}:1b), the explicit knowledge used for the matching calculations can be filtered by `Gender', `Age', `Body height' and `Body weight'. %represented and modified by changing the ranges of the individual STPs for different gait abnormalities and normal gait (see \figstyle~\ref{fig:prototypescreen1}:1b).
When selecting a category of interest, the loaded `Patient' can be compared to other patients in the `Parameter Explorer' view (see \figstyle~\ref{fig:prototypescreen1}:3). This table contains five columns. The first column represents the STP `ID' to create a connection to the number in the graphical summary represented in the formerly described `Knowledge Table'. In the second column, the `Parameter' name is represented and column three contains the calculated STP values for the loaded patient. The fourth column provides the `Interactive Twin Box Plot' (ITBP) \new{as shown in detail in} \figstyle~\ref{fig:interactivetwinboxplot}. \new{It is} an extended data visualization slider \cite{eick_1994_data} for inter-category comparison in relation to: 1) the `Norm Data category' represented as blue box plot; 2) the `Selected Category' of a specific gait patterns represented as orange box plot. A `Hatching Range Slider' (HRS) visualizing the related discrete parameters of each patient stored in the `Selected Category' and 3) the STP value of the currently loaded patient.
By placing the parts of the ITBP directly on top of each other, they can be perceived as a single control \cite{ware_2013_information}.
The ITBP enables the clinician to quickly compare two distributions and to set norm-value ranges for healthy and non-healthy categories. 
Additionally, based on the HRS, the clinician has the ability to quickly visually adjust the typical value ranges of the `Selected Category'.

The last column represents the difference $d$ between the `Norm Data Category' and the `Selected Category' which are visualized in the ITBP based on the Fisher discriminant function (see Equation~\ref{math:difference}) \cite{fisher1936use}:

\begin{equation}
	d = \frac{(\mu_{k}-\mu_{l})^2}{\sigma_k^2 + \sigma_l^2}
\label{math:difference}
\end{equation}

Hereby, for a given parameter, $\mu_k$ specifies the mean and $\sigma_k^2$  the variance of the first category $k$ and respectively $\mu_l$ specifies the mean and $\sigma_l^2$ the variance of the second category $l$. %In this context 
A %longer 
higher $d$ represents a larger difference between the parameter distributions of the two categories, yielding a wider bar.   
After a clinician has finished exploring the newly loaded patient data, he or she can add them to the currently selected knowledge table category in the EKS by using the `Apply' button. This way, the parameters of the patient are automatically transferred into the EKS,
recalculating the value ranges. 
Thereby, new explicit knowledge is generated and used for analysis support. Likewise, the clinician has the possibility to undo various changes in the EKS at any time by using the `Reset' button.

\subsubsection{EKS Exploration and Adjustment}
To support its second use case, \kavagait contains two additional views for the exploration and adjustment of the explicit knowledge stored in the EKS.
The clinician has the ability to select a single `Patient' in the EKS for comparison with other patients (see \figstyle~\ref{fig:prototypescreen2}:1). The ITBPs are showing the relation to the `Norm Data Category' and a `Selected Category' of abnormalities (see \figstyle~\ref{fig:prototypescreen2}:2). This visualization works the same way as formerly described for the exploration of newly loaded patient data \new{\sout{(see \figstyle~\ref{fig:prototypescreen1}:3)}}.
On the other hand, as shown in \figstyle~\ref{fig:prototypescreen3}:1, the clinician can select a category visualizing each STP value range set by HRS (see \figstyle~\ref{fig:prototypescreen3}:2) included in the `Selected Category'. Here, the clinician has the ability to change (overwrite) the automatically estimated range by moving the HRS for each parameter. This feature is needed for two specific cases: 1) If the EKS did not contain any patients in a category, the clinician has the ability to create the ranges for each STP of a category based on his or her implicit knowledge; 2) If an STP of a category contains outliers based on patient data, the clinician has the ability to readjust the range. Hereby, the color of the HRS will change to dark orange and by applying the changes, the category receives an orange triangle in the tree structure to remind the clinician that changes were applied by hand.
In general, each HRS included in the visual interface \new{\sout{(see \figstyle{}s~\ref{fig:prototypescreen1},~\ref{fig:prototypescreen2} and~\ref{fig:prototypescreen3})}} can be used for filtering or adjustment of the EKS included categories. The adjustments performed directly for the EKS will be stored permanently but they can be recalculated based on the values stored in the EKS if necessary. In contrast, the selected filtering options are automatically set back to default if the system is restarted.

\subsection{Interaction Concept}
For a better understanding of its functionality, we describe \kavagait according to five steps based on the visual information seeking mantra by Shneiderman~\cite{shneiderman_eyes_1996}:
overview first, rearrange and filter, details-on-demand, then extract and analyze further.

%\mypar{Overview}
\subsubsection{Overview}
When the clinician loads an input file, the patient's information and the ${F_{v}}$ data from the performed analysis are displayed in the center of the view \new{(see \figstyle~\ref{fig:prototypescreen1}%:2
)}.
The automatically calculated matching of the patient to the stored EKS categories will be presented on the left side\new{\sout{ (see \figstyle~\ref{fig:prototypescreen1}:1a)}}. Additionally, the `Graphical Summary' provides an overview of the 16 represented STPs including a comparison to the stored values.

%\mypar{Rearrange}
\subsubsection{Rearrange}
The clinician has the ability to rearrange each display, represented as a table, by sorting the columns\new{\sout{(see \figstyle{}s~\ref{fig:prototypescreen1}:1,~\ref{fig:prototypescreen1}:3 and \ref{fig:prototypescreen2}:2)}}.

%\mypar{Filter}
\subsubsection{Filter}
To reduce the number of patients used for the\sout{ calculation of the} automated category matching \new{calculation}, the interface offers the selection of several filtering options. Thereby, the clinician can filter the EKS data by `Gender', `Age', `Body height' and `Body mass'\new{\sout{(e.g., see \figstyle~\ref{fig:prototypescreen1}:1b)}}.
The matching results displayed in the `Knowledge Table' are updated immediately, and the graphical summary ('Parameters in Category') gives an impression of the 16 matched value ranges of the calculated STPs \new{(see \figstyle~\ref{fig:prototypescreen1}:1%a
).}

%\mypar{Details-on-Demand}
\subsubsection{Details-on-Demand}
If a matching result catches the clinician's interest, it can be selected from the knowledge table. 
This action opens a detailed visualization of the underlying parameters in a separate table -- the `Parameter Explorer' (see \figstyle~\ref{fig:prototypescreen1}:3),
which \new{can also be\sout{ also}} used for the exploration \new{of\sout{or} patients} already stored\new{\sout{ patients}} in the EKS (see \figstyle~\ref{fig:prototypescreen2}:2). In this table, the clinician can compare the calculated parameters of the loaded patient to the `Norm Data Category' and to the `Selected Category' from the `Interactive Twin Box Plots' (ITBPs).
%Additionally, 
Thus, the clinician still gets the information of how different the categories are for different STPs ('Category Difference').
It is important to note that the ITBPs are situated above each other to provide a better visual comparability of the differences between the \textit{left and the right foot}. This is important, as the clinician needs to assess differences between both body sides (gait asymmetry). Additionally, the clinician has the ability to sort the visualized data based on the different columns by clicking the respective header.

%\mypar{Extract}
\subsubsection{Extract}
Once the clinician has found the appropriate category for a patient's gait, the calculated parameters can be added to the `Selected Category' of the EKS by pressing the `Apply' button (see \figstyle~\ref{fig:prototypescreen1}:3).
Alternatively, the clinician can select some parameters of the patient in the `Parameter Explorer' table to add only them to the `Selected Category' of the EKS by using the `Apply' button.
From this moment, these data are immediately integrated into the automated analysis for the matching calculation. If a class contains insufficient samples or a value range is affected by outliers, the clinician has the possibility to extract further implicit knowledge by manually adapting these ranges in the `Category Parameter Explorer'. For this purpose, the ITBPs utilize the raw data of the selected class in order to provide the possibility for visual control by the clinicians (see \figstyle~\ref{fig:prototypescreen3}:2).

\subsection{Externalized Knowledge Integration}
% To support gait analysts during their work, we designed the EKS related to different specific gait patterns (functional anomalies) and norm data (healthy gaits). 
The EKS is included on the left side of the interface in two different forms depending on the task to be supported.

\subsubsection{\new{Usage of Knowledge}}
\new{\sout{First, w}}When the clinician explores newly loaded patient data, the EKS is presented in a table format (the `Knowledge Table' shown in \figstyle~\ref{fig:prototypescreen1}:1a).
All of the `Categories'\new{\sout{ which are}} integrated in the EKS will be checked against the loaded input data automatically.
Based on the explicit knowledge, the system distinguishes between the three states (in range, out of range, or no data) 
of the graphical summary for each of the 16 STPs in the `Parameters in Category' column.
Additionally, the system calculates how newly loaded patients match to the stored `Categories' in the EKS.
To add new knowledge to the EKS, the system provides two possibilities: 
On the one hand, the clinician can add the full patient dataset, representing each parameter as ITBP, by using the `Apply' button in the `Parameter Explorer' (see \figstyle~\ref{fig:prototypescreen1}:3) to the `Selected Category' in the `Knowledge Table'.
On the other hand, the user has the ability to select a set of parameters of interest from the `Parameter Explorer' table and add them using the `Apply' button in the `Parameter Explorer' to the `Selected Category'.

\subsubsection{\new{Adjustment of Knowledge}}
\new{\sout{Secondly, w}}When the clinician explores the explicit knowledge, and adjusts it for single patient data or a category, the EKS is presented as an indented list (the `Knowledge Tree') (see \figstyle~\ref{fig:prototypescreen2}:1).
On the one hand, \new{\sout{(see \figstyle~\ref{fig:prototypescreen2}:1)}} the clinician has the ability to select a single `Patient' from the EKS for comparison \new{\sout{(see \figstyle~\ref{fig:prototypescreen2}:2)}} with other patients by using the ITBP in relation to the `Norm Data Category' and the `Selected Category' including the selected `Patient' \new{(see \figstyle~\ref{fig:prototypescreen2})}.
On the other hand, \new{\sout{(see \figstyle~\ref{fig:prototypescreen3}:1)}} the clinician can select a category visualized by HRS \new{\sout{(see \figstyle~\ref{fig:prototypescreen3}:2)}} for each STP of the patients included in the `Selected Category' of the EKS \new{(see \figstyle~\ref{fig:prototypescreen3})}.
Generally, at the end of each `Category', the number of contained `Patients' is shown in blue brackets.

\begin{figure}[t!]
	\centering
		\includegraphics[width=1.0\columnwidth]{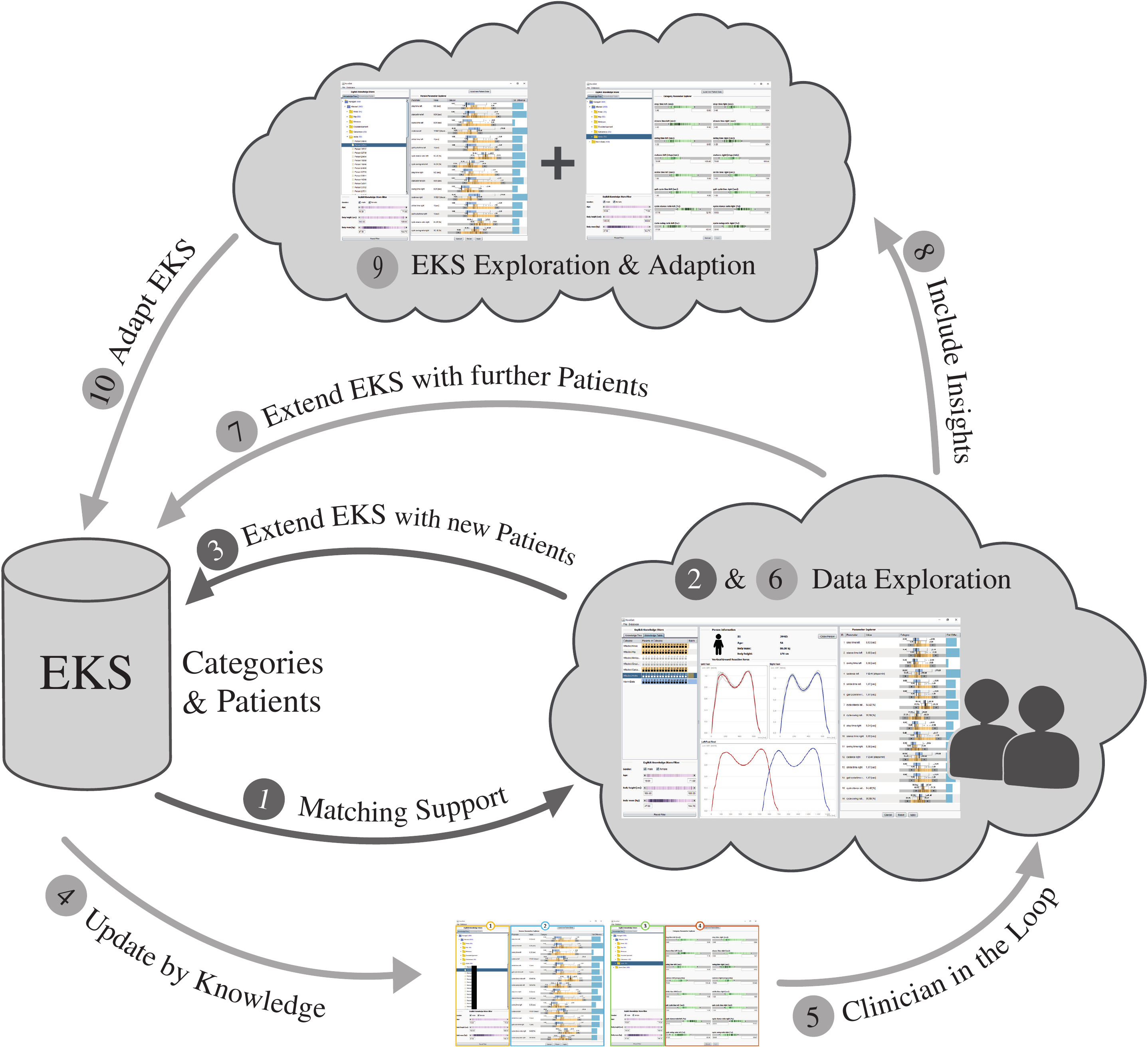}
      
		\caption{Overview of the system's knowledge generation loop, beginning with the dark gray inner loop after loading a new patient and changing to the light gray outer loops for interactive data exploration and knowledge generation. The clinician plays a major role in both loops.}
	\label{fig:knowledgeloop}
\end{figure}
  
%\mypar{Knowledge Generation Loop}
\subsubsection{\new{Knowledge Generation Loop}}
\figstyle~\ref{fig:knowledgeloop} provides an overview of the system's knowledge generation loop, starting at the dark gray inner loop. 
In general, the system's EKS stores all `Patient' data in several `Categories' depending on patients' pathologies (gait abnormalities) which were generated by former gait analysis sessions.
If the clinician loads a new patient file, the calculated STPs will be checked automatically against the EKS (1) to calculate the category matching.
Depending on the automated matching calculations, the system provides a visual representation of the results (2).
From this point, the clinician can carry out the patient data exploration and analysis, thus, the clinician is an important part of the knowledge generation loop.
(3) During the patient data analysis driven by the clinician, the clinician has the ability to include the patient data into a `Category' in the EKS.
By adding a new `Patient' to the EKS or setting filters, the system automatically refreshes the matching calculations depending on the explicit knowledge stored in the system (4). This brings the user into the outer (light gray) part of knowledge generation loop.
Here the clinician is part of the continuously recurring loop (5), for data exploration (6) and knowledge generation (7). 
Additionally, the clinician has the ability to continuously include new insights (i.e., gait categorizations, value range limits) (8) depending on `Patient' and `Category' exploration and adjustment (9) to adapt the EKS value ranges, which are predefined by the stored explicit knowledge, for further automated analysis (10).

%%%%%%%%%%%%%%%%%%%%%%%%%%%%%%%%%%%%%%%%%%%%%%%%%%%%%%%

% Only needed for my PhD work (MW)
%\section{Functional Overview}
%\label{sec:functionaloverview}
%\input{Chapters/5_FunctionalOverview.tex}

%%%%%%%%%%%%%%%%%%%%%%%%%%%%%%%%%%%%%%%%%%%%%%%%%%%%%%%

\section{Validation \& Results}
\label{sec:validationandresults}
To validate the \kavagait system and provide evidence for its effectiveness, we followed a threefold research approach consisting of moderated expert reviews, user studies~\cite{lazar_research_2010} and a case study with a national expert. 
All of the insights were documented in detail to ensure reproducibility~\cite{smuc_should_2015} and used to improve our research prototype.
All materials used, such as interview guidelines and tasks, are available as supplemental material \suppmat.

%%%%%%%%%%%%%%%%%%%%%%%%%%%%%%%%%%%%%%%%%%%%%%%%%%%%%%%%%%%%%%%%%%%

\subsection{Expert Reviews}
In the first step, we conducted iterative expert reviews to eliminate usability issues in the basic functionality and appearance of the interface. 

\subsubsection{Method}
\mypar{Participants}
To validate the visual interface design, we invited three usability experts. %(see \tabstyle~\ref{tab:usabilityexperts}). 
Each of them has between two to four years of experience in this field.
Two of them are between 20 and 29 years of age and one is between 30 and 39 years of age. 
All of them have a Master's degree and advanced or expert knowledge in usability. 

\mypar{Design and Procedure}
Each usability expert received a short introduction to the basic features and the workflow of the system.
Next, each expert  walked through each feature individually and assessed usability issues. 

\mypar{Apparatus and Materials}
As evaluation material, we generated builds of \kavagait in different development states and used them for the iterative expert review sessions performed on a 15$''$ notebook with a full HD screen. 
%The review sessions were performed on a 15$''$ notebook with a full HD screen resolution and an external mouse for navigation.  
Each expert review was documented in short notes on paper by the investigator.
%\done{ich nehma mal an, dass dieser Teil hier mit redundanz gemeit war, habs mit (1) gekennzeichnet unterhalb in meinem Kommentar um die Zugehörigkeit zu sehen.}

\subsubsection{Results}
The basic color scheme of \kavagait was found to be easily recognizable. 
Only the coloring of the 'Categories' was pointed out as being not well differentiated from the other elements (see \figstyle~\ref{fig:prototypescreen1}). 
The visualization metaphors (boxes, folders and sheets) for the knowledge tree visualization was developed in conjunction with the usability experts to represent a familiar structure to the analysts. 
The experts suggested that it is necessary that the interface automatically applies the entered parameter if the user left the focus of a filtering input box. 
Overall, all of the usability experts provided positive feedback on the design structure of the system.   
All of the expert's suggestions were used for a redesign and revision of the system in order to prevent the domain users from having basic interface issues. 

%%%%%%%%%%%%%%%%%%%%%%%%%%%%%%%%%%%%%%%%%%%%%%%%%%%%%%%%%%%%%%%%%%%

\subsection{User Study}
A user study with six gait analysis experts was performed in October 2016 as formative evaluation of usability~\cite{cooper_about_2007} on the revised system.
Each test took approximately 1.5 hours and encompassed four analysis tasks, the system usability scale questionnaire (SUS)~\cite{brooke_sus_1996}, and a semi-structured interview built upon 13 main questions to be answered.
The user study's goals (G) and non-goals (NG) are defined as:
(G1) Testing the functionality of the research prototype;  
(G2) Testing the visualization techniques for comprehensibility in relation to the domain;
(G3) Testing the utility of knowledge storage and representation in the system;
(NG1) Comparison of \kavagait with another analysis system and
(NG2) conducting performance tests, because there was no comparable interactive analysis system found for this domain.   

\subsubsection{Method}
\mypar{Participants}
We invited six gait analysis experts (see \tabstyle~\ref{tab:participants}) to participate in the user study. 
One participant had given feedback on sketches and early prototypes previously as a member of the focus group for the user-centered design process
(see \secstyle~\ref{sec:designandimplementation}).
All experts work in the field of clinical gait analysis as physical therapists,
bio-medical engineers, or sports scientists.
Additionally, all of them are involved in different gait analysis or physical therapy research projects and two of them are also working as physical therapist in a hospital. 

\begin{table}[ht!]
		\caption{Data on user study participants describing  education, years in the field and knowledge in gait analysis. (Gender: f~$:=$~female, m~$:=$~male; Organization: R~$:=$~research, F~$:=$~faculty, H~$:=$~hospital; Knowledge: 1~$:=$~basic, 2~$:=$~skilled, 3~$:=$~advanced, 4~$:=$~expert)}
		\label{tab:participants}
	\resizebox{1.0\columnwidth}{!}{\begin{minipage}{\columnwidth}
		\centering
        \begin{tabular}{ p{1.2cm} p{1.0cm} p{1.0cm} p{1.0cm} p{1.05cm} p{1.0cm}}\hline
			\tablefirstrowformat{Person (Gender)} & 
			\tablefirstrowformat{Organi\-zation} & 
			\tablefirstrowformat{Age} & 
			\tablefirstrowformat{Know\-ledge} &
            \tablefirstrowformat{Years in Field} &
			\tablefirstrowformat{Edu\-cation} \\ \hline
			\rowcolor{gray!15}
            \tabletextsize{P1 (f)}  & \tabletextsize{F}    & \tabletextsize{40-49} & \tabletextsize{2} & \tabletextsize{15}  & \tabletextsize{MSc} \\
			\tabletextsize{P2 (m)}  & \tabletextsize{F}    & \tabletextsize{40-49} & \tabletextsize{3} & \tabletextsize{15}  & \tabletextsize{PhD} \\
			\rowcolor{gray!15}
            \tabletextsize{P3 (m)}  & \tabletextsize{R\&F} & \tabletextsize{30-39} & \tabletextsize{2} & \tabletextsize{5}   & \tabletextsize{PhD} \\
			\tabletextsize{P4 (f)}  & \tabletextsize{R}    & \tabletextsize{40-49} & \tabletextsize{1} & \tabletextsize{1.5} & \tabletextsize{MSc} \\
			\rowcolor{gray!15}
            \tabletextsize{P5 (f)}  & \tabletextsize{F\&H} & \tabletextsize{30-39} & \tabletextsize{2} & \tabletextsize{10}  & \tabletextsize{MSc} \\
			\tabletextsize{P6 (f)}  & \tabletextsize{F\&H} & \tabletextsize{20-29} & \tabletextsize{2} & \tabletextsize{7}   & \tabletextsize{MSc} \\ \hline
		\end{tabular}
	\end{minipage}}
\end{table}

\mypar{Design and Procedure}
At the beginning of the user study, each participant was asked about a general impression of the user interface and which functions could be recognized. 
This step took approximately five minutes.
Subsequently, each participant had to solve \textbf{four guided analysis tasks}. 
The first three included a step-wise introduction to the system and the last one was a combined analysis task to determine the understanding of the system's workflow (this step was also required for the subsequent SUS questionnaire). 
Each analysis task was read to the participant at the beginning of the task, and for reference, each task was handed over to the participant in printed form.
For the analysis tasks, participants spent approximately 40 minutes. 
After the analysis task session, each participant had to fill out a standardized \textbf{SUS} questionnaire~\cite{brooke_sus_1996} in less than five minutes. 
%The SUS is a standardized, technology-independent questionnaire to evaluate the usability of a system~\cite{brooke_sus_1996}. 
%During this test, the participants were asked ten questions on a five-level Likert scale from strongly agree to strongly disagree. 
%
Finally, we performed \textbf{semi-structured interview} sessions with an average duration of 40 minutes.
For this, we used an interview guideline consisting of 13 major questions addressing general system usability, filtering, using the `explicit knowledge store' (EKS),  
and individual visual metaphors used in \kavagaitshort.

\mypar{Apparatus and Materials}
The user studies were performed in a silent room and the \textbf{analysis tasks} were performed on a 15$''$ notebook with full HD screen resolution and an external mouse.  
%We used a silent and clean room without any distractions to perform the three parts of the user study for all participants under the same conditions.
%The four \textbf{analysis tasks} were performed on a 15$''$ notebook with full HD screen resolution and an external mouse.
As datasets for the analysis tasks, we used two anonymous clinical gait analysis samples recorded by an AUVA clinical gait laboratory. The provided datasets included one healthy and one patient with a gait abnormality. 
To achieve the best outcome, we asked the participants to apply thinking aloud~\cite{nielsen_usability_1993} during the whole analysis task part.
For further analysis of the user test, we recorded the screen and the participant using the notebook's internal webcam. 
In parallel, the facilitator took notes in our pre-defined test guideline.
The \textbf{SUS} questionnaire and \textbf{semi-structured interview} were conducted on paper in the participants' native language.
For the detailed questions, we used small images in the semi-structured interview guidelines to support the participants in recalling the respective parts of \kavagaitshort{}.

\subsubsection{Results}
The following section describes the major findings of each part of the test and summarizes their results.

\mypar{Analysis Tasks}
During the four analysis tasks, all participants described their solutions to the tasks at hand by thinking aloud. 
They expressed problems as well as benefits of \kavagait during their work. 
One major problem during the test was that the participants tried to find out which rectangle in the graphical summary, displayed in the knowledge table, relates to which `Interactive Twin Box Plot' (ITBP) in the parameter explorer [P2, P3, P5, P6].
In relation to the graphical summary, all participants understood the gray rectangle, the empty black frame and the black rectangle shape encoding the matching of the individual parameters in relation to 'no data available’, 'out of range’ and 'is in range’. 
%Three out of six participants suggested that the coloring of the box plots while comparing norm data with norm data should change to blue for both contained box plots.
Four out of six participants stated during the third analysis task that they confirm the automated matching result 
after comparing to the vertical ground reaction force ($F_v$) data.
In general, all participants stated 
that the automated matching can be used as a guided entry point for further analysis using the twin box plots to analyze the patients’ data in detail. 
Additionally, all participants handled the tasks depending on the knowledge exploration or own knowledge externalization very well. 
They understood the knowledge tree organization metaphors (boxes, folders and sheets) visualizing categories and individual persons. 

\mypar{System Usability Scale (SUS)}
By applying the SUS, we were able to obtain a good indication concerning the handling of our \kavagait system.
The results show a SUS value of 80 points in average out of 100, which can be interpreted as good without significant usability issues according to the SUS description by Bangor et al.~\cite{bangor_determining_2009}. 
Comparing 500 SUS scores, Sauro~\cite{sauro_measuring_2011} found that only 10\% of systems reached an SUS score greater than 80.

\mypar{Semi-structured Interviews}
Next, we present the results of the performed interviews, which were structured along 13 main questions.
%All participant statements quoted in this section have been translated from their native language %German 
%to English by the authors. 
 The detailed evaluation of the interviews can be found in our supplement material \suppmat.

All participants attending the user study confirmed a very clear system design. 
The used visualization metaphors are described as well interpretable, the color scheme is conclusive and the calculated matching acts trustworthy. 
The integrated filter options for data reduction make sense and are understandable. 
%However, it was questioned whether clinicians should use such filtering options, as these may bias their interpretation. 
The various visualization options included in \kavagait and the curves and box plots contributed very well and were considered understandable. 
% Only one participant was confused because of her missing statistical background knowledge.
Only one participant, P6, indicated that it was not clear how the 'Category Difference' was calculated. %and she lacked statistical background knowledge to comprehend Equation~\ref{math:difference}.

Generally, all participants told us that they readily understood the EKS, including the ability to compare newly loaded patients to the EKS for categorization.
Additionally, a single system would be very beneficial as well as a shared EKS. 
The knowledge representation options ('Knowledge Tree’, 'Knowledge Table’, `Hatching Range Slider' (HRS) and ITBP) were described as very helpful and good for quick decision making. 
They described that it was easy to get an overview of the loaded patient in relation to the 'Norm Data Category’ and the 'Selected Category’ based on the ITBPs. 
Thus, the analyst has the ability to navigate through the data or to rely on the EKS based matches. 
Additionally, the participants referred to the categories of the EKS, including patient data of prior analysis, as well as to the different representations in the 'Knowledge Table’. 
Four out of six participants reported that based on the ITBP the sharp distinction, interval range, variance, differences and relationships can be derived for the analysis.

All participants noted that they have understood the saving process to the EKS. 
The category in which a patient is stored should appear subsequently. 
As 'nice to haves’, a separated saving area for not yet diagnosed patients and the ability for annotations would be helpful. 
The separated category should not be included in the automated matching calculations. 
Furthermore, all participants reported that they understood the symbols represented in the 'Knowledge Tree’ of the EKS but most of them did not assign any further meaning to them. 
Based on the different colors it was easy for the participants to distinguish the different EKS levels. % of the EKS.

In relation to the 'Knowledge Table’, the participants classified the 'Graphical Summary’ as very helpful. 
Only two participants had small issues at the beginning when interpreting this visualization metaphor. 
Based on the coloring and the boxes for the parameters it was easy to understand the matching results. 
The matching criterion was indicated as helpful and good starting and orientation point for analysis. 
Some participants argued that a connection based on sequence numbers between the 'Graphical Summary’ and the ITBPs would be very helpful. 
This suggestion and other suggestions were subsequently implemented as shown in \figstyle{}~\ref{fig:prototypescreen1}.
The range slider shading and its usability for range assessment was quickly recognized by most of the participants (e.g., used for filtering, category parameter exploration and adjustment). 
The shading gave an overview of the underlying data derivation and outliers. Additionally, it helps to explain the ITBPs shape.

All participants had prior exposure to statistical box plots and therefore readily understood the ITBPs.
They noted that this metaphor contains a lot of information but it is clearly structured. 
Additionally, some participants stated that the usage of the same color will be better understandable when selecting the 'Norm Data Category’ for comparison in the ITBP (see \secstyle~\ref{sec:designandimplementation}) in the future. %Angepasst nach Aussage von R2 bei den Rechtschreibfehlern
The comparison of a (newly loaded) patient with the 'Norm Data Category’ and a 'Selected Category’ based on the ITBPs was described as helpful. 
It is possible to see the differences of the individual category parameter, but you have to know how to interpret them.

In general, the \kavagait system was described as very innovative and helpful for the analysts by providing automated analysis and pointing out possible reasons to be respected in clinical decision making.

%%%%%%%%%%%%%%%%%%%%%%%%%%%%%%%%%%%%%%%%%%%%%%%%%%%%%%%

\subsection{Case Study}
\label{sec:casestudy}
After finalizing the expert review and the user study we made the following improvements before conducting the final case study: 1) ID numbers for parameter identification were added to the ITBP and the `Graphical Summary'; 2) The coloring scheme of the ITBP was improved; 3) Tool tips were added to all elements; 4) The entire labeling used in the system was checked for consistency. 

\subsubsection{Method}
\mypar{Participants}
For our case study, we invited one leading national expert for gait rehabilitation to test and comment on our novel \kavagait system. The expert has more than one decade of experience in conducting clinical gait analysis. Thus, the expert is comfortable in identifying gait patterns based on the representation of ground reaction forces (GRFs) and the calculated spatio-temporal parameters (STPs).

\mypar{Design and Procedure}
At first, the expert received a short introduction, in the form of a presentation, about the basic features and workflow of the system. Next, the expert walked through each feature individually by an example and was asked to critically comment on the system. Additionally, the expert could choose different patients from our data pool to explore them and tell us new insights gained using \kavagaitshort.

\mypar{Apparatus and Materials}
We met in the expert’s office room to perform our case study. As materials, we used a short presentation of the \kavagait system and a build of the revised prototype including 489 anonymized patients in the `Norm Data Category’ and 50 patients for each out of four patient categories (ankle, calcaneus, hip and knee associated gait abnormalities). The case study was performed in the same setup as in the user study before. 
The suggestions and comments stated by the expert were documented  by the presenter and one observer.

\begin{figure}[t!]
    \centering
    \begin{subfigure}[b]{1\columnwidth}
        \includegraphics[width=\columnwidth]{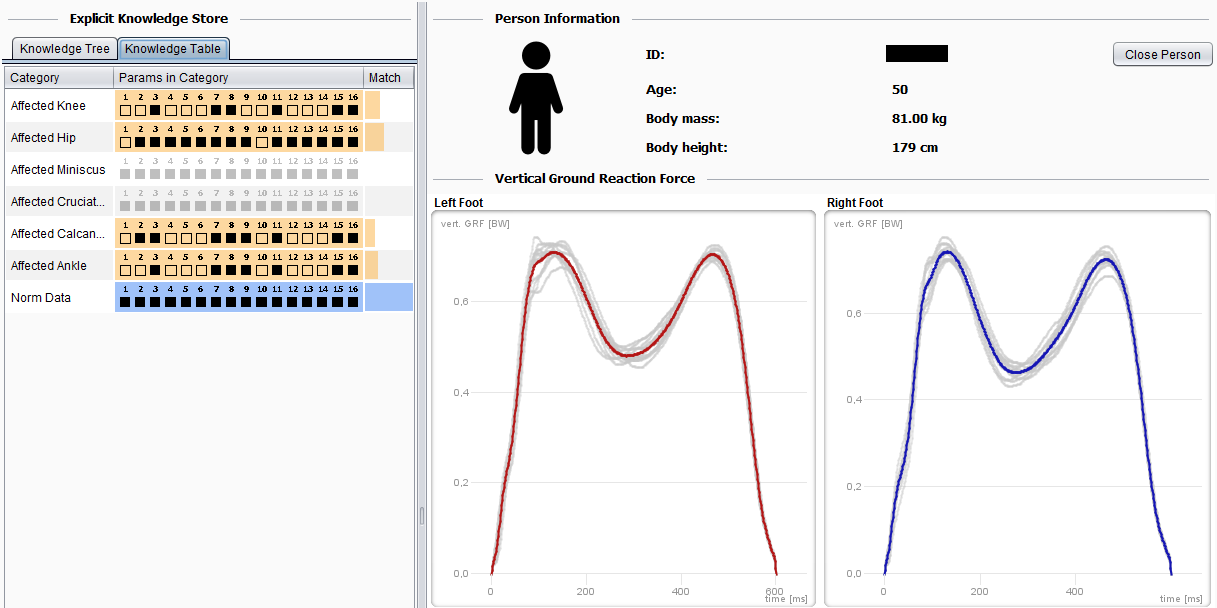}
        \caption{Normal gait}
        \label{fig:normalgait}
    \end{subfigure}
    \begin{subfigure}[b]{1\columnwidth}
        \includegraphics[width=\columnwidth]{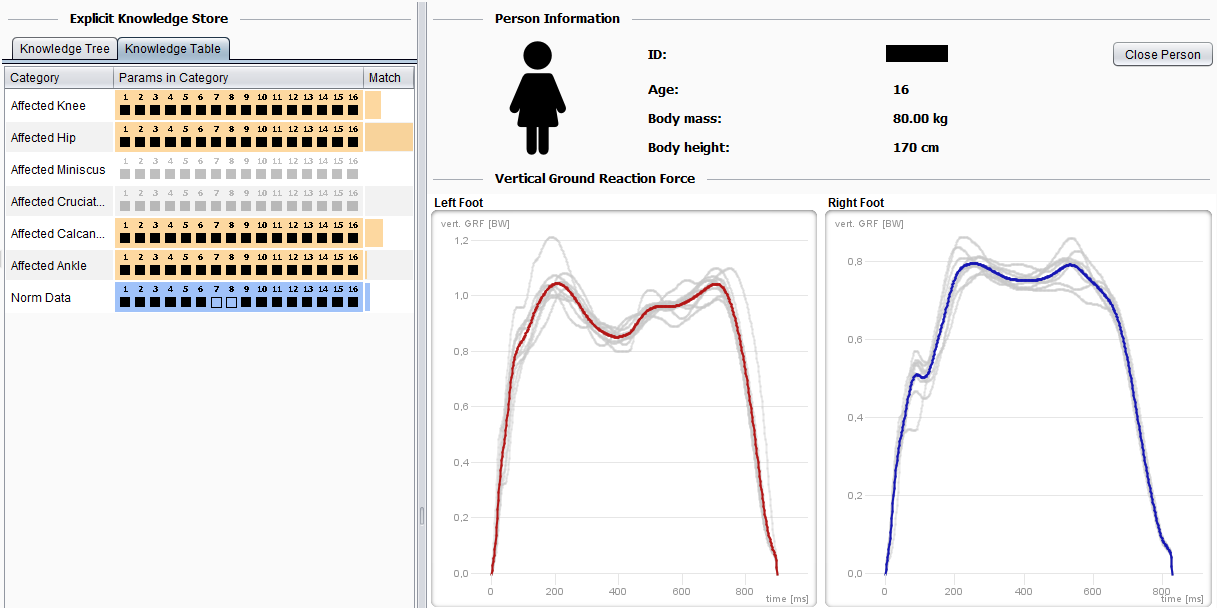}
        \caption{Hip associated gait abnormality}
        \label{fig:hipabnormality1}
    \end{subfigure}
    \caption{Examples of the gait analysis data sets presented and discussed during the case study containing patterns of (a) normal gait and (b) gait abnormality.}
    \label{fig:casestudyexamples}
\end{figure}

\subsubsection{Results}
The expert initially noted that a clinician normally focuses on two major aspects. Firstly, they look for asymmetries comparing the STPs between the left and the right foot and secondly for deviations of them %with respect 
regarding to the norm data. %which are collected by performing thousands of clinical gait analyses over the years. 
%The walking speed of the patients is not of interest because it depends on the day time and their constitution. 

\mypar{EKS Patient Exploration} 
\new{\sout{Next,}} The expert randomly selected a patient stored in the `Norm Data Category’. %This category normally includes only the best trials of a patient (used view see \figstyle~\ref{fig:prototypescreen2}:2). 
%For a fast comparison, the expert would prefer an initial sorted representation of the ITBPs by parameters (e.g., Stance time left/right, Swing time left/right). 
Generally, the expert found an added value in relation to the ITBP representation for the STPs. A key statement by the expert was: ``You do not have to hit the median value directly, it is more important if the parameter values for the left and the right foot are looking similar'' near to the median value of the `Norm Data Category'. 
By randomly selecting a patient from the `Knee Category’ for exploration, the expert was immediately able to sort the STPs. 
%By orienting on the mean values represented from the `Norm Data Category’ and the `Knee Category’ the expert stated that the values did not look so bad, most of the patient values also matches with the `Norm Data Category’.
An additional suggestion was to provide the ability to select two groups for comparison. In this case, for example, the expert could compare the `Knee Category’ to the `Ankle Category’ without changing by viewing the related mean curves of each category with their associated one-standard deviation bands. Additionally, a more detailed separation of the datasets (e.g., for abnormalities in the left, the right or both feet) in the EKS was stated as helpful to activate and deactivate these groups directly in the future.

\mypar{New Patient Data Exploration} To test the analysis abilities of \kavagaitshort, the expert successively loaded four clinical gait analysis records, which were containing gait abnormalities (1x norm, 2x hip, 1x ankle). Two of these examples are illustrated in \figstyle~\ref{fig:casestudyexamples}. %The loaded analysis files contain gait abnormalities on the left, the right or both sides. 
The representation of the individual steps represented as light gray curves in the background with their corresponding mean value curve was described as very helpful. The expert analyzed each loaded patient file exactly with regard to 
the $F_v$, the calculated parameters represented in ITBPs, and the system-internal computed matching based on the EKS.
During the case study, several of the assessed records indicated multiple matching to different gait patterns (which were reported in the `Knowledge Table'). The expert was able to confirm the EKS based matching results, and reported that the system was of valuable support during gait record examination.
This can be attributed to the fact that the patients usually undergo a therapy for a specific problem. However, \kavagait also recognizes abnormalities in other joints caused by specific problems. The expert stated that the community has always been aware that one abnormality can lead to others, but it has never been so well presented. %, as in \kavagaitshort.

\mypar{Concluding Discussion} To date the system only incorporates a total of approximately 500 patients. Thus, in each category the chance for imbalances between sub-groups, such as defined by gender, age, body mass, or height may be present. These in consequence, may introduce a bias when calculating the final matching. Future prospects may include a direct connection the AUVA database to include a sufficient and representative number of patients per class to overcome this problem.

The expert also noted, that a further development could be to visualize and store additionally derived and commonly used discrete parameters (next to the STPs) from the entire GRF curves, such as local minima and maxima, loading rates, among many others. This would further strengthen the system's capabilities of describing a patient's gait performance. Another feature, noted to be of valuable interest, would be the possibility to compare a patient with earlier treatments.
This would help to visualize the entire rehabilitation process over time.

In general, the expert outlined that the system supports the process of gait assessment in several ways: 1) %In our data we observed a bias towards gender-unbalanced gait data. This is evident when filtering by gender. The body mass index, the body height or the body mass should normally not influence the stored data for comparison if the EKS contains enough data (e.g., \new{future} connection to the AUVA database). 
The system easily allows to compare new patient data to stored EKS data, and thus helps to gather a more conclusive picture of the patient's gait performance, which was not possible in clinical practice before.
2) The matching criteria of \kavagait helps to clearly visualize and identify secondary gait impairments. This is very important, as during clinical examination secondary gait impairments are easily overlooked. 3) %The visual representation that one abnormality can lead to others, which can be seen in the matching criteria;
The ability to compare the norm data with a selected category for direct comparison based on the ITBP is very helpful to gain further insights into the patient's gait performance and possible impairments.
4) \new{In addition, the \kavagait system is also well suited for educational training of unexperienced clinicians.  % as well as for analysts who do not perform clinical gait analysis every day. 
%For experts who are working every day in this area, the system can be a good supplement during the analysis process.
Overall, the expert described \kavagait as an excellent and helpful analysis tool for clinical practice.}

%%%%%%%%%%%%%%%%%%%%%%%%%%%%%%%%%%%%%%%%%%%%%%%%%%%%%%%

\section{Reflection \& Conclusion}
\label{sec:reflectionandconclusion}
Following the design study methodology of Sedlmair et al.~\cite{sedlmair_design_2012}, the reflection (including retrospective analysis and lessons learned) is the third contribution of a design study which enables the improvement of current guidelines.
The following paragraphs describe the reflection in line with the initially stated requirements from \secstyle~\ref{subsec:prototyperequirements} (R1 -- R4). %The different validation steps confirmed that our \kavagait prototype fulfills the requirements for clinical gait analysis. %MZ: der Satz passt hier nicht so gut

\question{R1 Data:}
The data structure resulting from force plate recordings, are synchronized time series data of the  ground reaction forces (GRF).  
Especially in a clinical gait analysis setting, these data comprise time series of two force plates (one per foot). Based on these data, we calculated \new{16\sout{eight}} additional STPs (spatio-time parameters) parameters, which were used for automated and visual comparison.
%Based on these calculations, the clinicians get the ability to gain new insights by comparing a single patient to different datasets (a 'Norm Data’ category or specific gait pattern categories).
On the one hand, we designed the `Graphical Summary' for data comparison, showing the analyst how a single patient parameter is related to the parameter set of a category. 
On the other hand, we designed the `Interactive Twin Box Plot' (ITBP) for detailed inter-category parameter comparison between the `Norm Data Category', a `Selected Category' of a specific gait pattern and the patient to analyze. 
%In clinical gait analysis typically additional discrete parameters are calculated from the entire curves (e.g. loading rates, etc.) to give a more detailed picture of a patient's gait performance. A future prospect should be to include such parameters as well clinics~\cite{benedetti_data_1998}.
%Another challenging future direction is to integrate machine learning for improved and automated patient categorization based on unsupervised (e.g.,~\cite{nuesch2012gait,christian2016computer}) and supervised (e.g.,~\cite{pauk2016gait,sangeux2015sagittal}) approaches. Both aspects will support in drawing more precise medical decisions based on the data available.

\question{R2 Visual Representation:}
In general, the decision for an interface providing a clear structure and understandable visual representations was well received.
It was easy for the domain experts in our validation to understand the handling and to work with it. 
Additionally, they appreciated the prototype's wide range of coordinated features, which they regarded as useful for data exploration and analysis while not being overloaded. 
A particularly interesting outcome of the tests from the visualization design perspective was, that the HRS are very useful for a parameter overview and range adjustments during patient data exploration as well as for the `explicit knowledge store' (EKS) exploration and adaption. 
Additionally, the ITBPs were considered as well-suited for intercategory comparison in relation to a single patient parameter. This way, it is possible to see how well the patient develops in the direction of the `Norm Data Category' for example. 
Another particularly notable outcome is that the `Graphical Summary' and the `Matching Criteria', represented in the `Knowledge Table', were described as very valuable by the national gait rehabilitation expert. 
%Seeing how a specific gait pattern affects other abnormalities is a very helpful insight, which can be discovered easily with the \kavagait system but was not brought to attention in the current setting.
To date the system only provides information of the vertical GRF component. To increase the quality of the analysis, in the future we will extend the visualization to all three force components of the GRFs.
In addition, we will also add discrete key GRF parameters of  those components to further improve automated analysis and exploration.
Additionally, direct comparisons of left and right foot (i.e., gait asymmetry indices~\cite{cabral2016global}) need to be integrated in the twin box plots.

\question{R3 Workflow:}
All included filter methods and the dynamic query concept providing a very fast response in \kavagait were very well received. 
In general, the participants described the filtering, analysis, and exploration abilities as intuitive and the usage and adaption of the EKS as easy to use.
%As further improvement before the case study, we added an ID number for each of the 16 STP to easier relate between the 'Graphical Summary’ and the ITBPs depending on the sorting abilities in the 'Parameter Exploration’ view. 
Our national expert mentioned that \kavagait is very valuable and suits several use cases, such as support for clinical experts, assistance for less experienced clinicians, and learning and training opportunities for students. 
Based on the insights we gained during our validation studies we found that for the participants and the national expert, the visual representations of the expert knowledge and the handling of the EKS was simple to understand and to use.
To further improve the workflow, the ability to annotate the patients' data would be a helpful feature. 

\question{R4 Expert Knowledge:}
As previously mentioned, the `Knowledge Tree' and the `Knowledge Table' of the EKS were well received by the participants and the case study member. 
The knowledge organization as boxes, folders (categories), and sheets (patients) was well received by most of the participants. 
Based on the counter after each category (see \figstyle{}s~\ref{fig:prototypescreen1}:1 and~\ref{fig:prototypescreen1}:3), it was easy to understand how meaningful the data are for comparison and to get an overview of the included data.
%Reflecting on the insights gained in the validation process, it can be concluded that a clinician can benefit from the EKS. 
%we found out that the analysts appreciated and could benefit from the externalized expert knowledge by sharing and adapting EKS elements during the analysis process.
In future explicit knowledge should also be used in the VA workflow to train machine learning methods for improved  automated categorization. 

%\mypar{Lessons Learned}
\subsection{\new{Lessons Learned}}
As described in \secstyle~\ref{sec:backgroundandproblemcharacterization}, clinicians currently are using non-interactive line plots and tables. For clinical decision making they are using their implicit knowledge based on several years of experience.
During this design study, we learned that explicit knowledge extracted from the clinicians implicit knowledge opens the possibility to support clinicians during clinical decision making. Additionally, \kavagait could also be used to share the knowledge of domain experts and for educational support. 

For keeping up with the large number of patients stored in the EKS, clinical gait analysts need to continuously adapt the systems settings during the clinical decision making process.
Supporting such interactive workflows is a key strength of visualization systems. 
Clinical gait analysis in particular profits from extensive interaction and annotation because it is a very knowledge-intensive job. 
By providing knowledge-oriented interactions, externalized knowledge can subsequently be used in the analysis process to support the clinicians.
Our newly developed visual metaphors provide an easy way to inspect variability of the data (e.g., standard deviation), allow to identify outliers in the data, and provide an easy to understand overview of the data and automated matching results (as demonstrated in \figstyle~\ref{fig:prototypescreen1}:1a).   
Additionally, based on the ITBPs \new{\sout{(see \figstyle~\ref{fig:interactivetwinboxplot})}} it is possible to perform intercategory and patient comparisons by details on demand to find similarities in the data. % and to gain new insights. 

%\mypar{Limitations \& Future Directions}
\subsection{\new{Limitations \& Future Directions}}
\kavagait is a design study investigating how interactive knowledge-assisted VA methods can aid clinical decision making in the context of gait analysis. Since the system is still a proof of concept, some limitations exist. These, however, point out future  directions of research in both areas, VA and clinical gait analysis. Currently, the proposed system only incorporates the vertical ground reaction force component, as used by several studies \cite{muniz2009application,lozano2010human}.
Nevertheless, it is subject to future work to include the other force components as well. This will help the clinicians to get a more holistic view of one's gait performance and will further strengthen the system's capability in supporting clinical practice. Another limitation might be associated with the defined parameters for matching and comparing purpose.
To date only a set of the most commonly used STPs are included.
However, there are several other discrete parameters that could be used for analysis and matching purpose.
Research has shown that sophisticated machine learning algorithms bear the potential to identify and cluster gait patterns \cite{soares_principal_2016,zeng_parkinsons_2016}. 
For example automated patient categorization based on unsupervised (e.g.,~\cite{christian2016computer}) and supervised (e.g.,~\cite{pauk2016gait}) approaches might be interesting. Both aspects will support in drawing more precise medical decisions based on the data available.
Results, however, clearly state, that the entire waveforms as input variables result in higher classification accuracies than using discrete parameterization techniques. Thus, future work might opt to include both, discrete parameters to inform clinicians and machine learning techniques, which use the entire waveforms, to allow for more advanced pattern recognition abilities and classification functionalities.  
Another future direction might be to provide the ability for searching the most similar dataset stored in the EKS to the currently viewed patient data. At this time, such a mechanism is not included in \kavagaitshort. Currently, \kavagait offers the ability for comparison of each parameter for both feet of one patient. Gait symmetry plays a key role for clinicians in analyzing and interpreting gait data. To date \kavagait only allows for visual inspection of parameters of the left and right body side. To increase the quality of such comparisons, parameters such as the `gait asymmetry index’ (GAI)  might be valuable and should be included in future versions (e.g., \cite{herzog_asymmetries_1989}). %In general, it is necessary to include further improvements into the \kavagait in combination with further comparative evaluations.
Finally, the clinical decision support provided by KAVAGait needs to be evaluated for the effect of cognitive biases such as the confirmation bias \cite{nickerson_confirmation_1998} that might increase due to previously externalized knowledge and interactive steering. 
KAVAGait addresses such concerns by prominently providing clinicans with raw GRFs displayed as curves. Our interviews indicate that these GRF curves are always considered in decision making.
A further research direction is also the integration of information on provenance and certainty into the EKS.

%\mypar{Transferability}
\subsection{\new{Comparison to the State-of-the-Art}}
\new{From a visual analytics perspective, this is the first design study in clinical gait analysis.
Related work on human motion analysis \cite{vogele_efficient_2014, bernard_motionexplorer:_2013, jang2016} focuses on clustering of motion segments recorded by a tracking system.
The gait analysis systems used in clinical practice typically come with gait analysis hardware
and present the collected data of only one measurement session (of one patient) in a non-interactive interface, as line plots in combination with several calculated discrete parameters.}
In contrast to \new{those,
\kavagait enables the comparison of a patient's measurements with an entire knowledge store of pre-categorized measurement sessions (potentially thousands of previous measurements),
interactive filtering, and the creation of new knowledge. Hence, a direct comparison between \kavagait and currently available systems would be limited in terms of validity.}
\kavagait uses analytical and visual representation methods to provide a scalable and problem-tailored visualization solution following the VA agenda~\cite{thomas_illuminating_2005, keim_mastering_2010} and none of the existing systems provides a comparable basis.

\subsection{\new{Transferability \& Generalization}}
The knowledge generation loop (see \figstyle~\ref{fig:knowledgeloop}) can be generalized for other domains taking into account domain-specific data structures and patterns of interest. 
\new{In the recently published ``Conceptual Model of Knowledge-Assisted Visual Analytics''
by Federico and Wagner et al.~\cite{federico_knowledge_2017}\new{,}
KAVAGait serves case study of how explicit knowledge is integrated in VA workflows.
On a general level, the workflows for knowledge generation and extraction 
always \sout{include the user as an integral part of the loop} % \cite{endert_2014_human}.
need to involve the human expert.
The methods to integrate knowledge and their applicability, generally depend on the type of the underlying data and the way how the explicit knowledge can be stored and connected to data.
In context of \kavagait, we focused on easy to understand parameter summarizations (`Graphical Summary') including automated matchings, the HRS \sout{(see \figstyle~\ref{fig:prototypescreen3}:2)} for deviation and outliers observation and the ITBP for the comparison of different categories whereby each ITBP is used for one parameter set.}

\ifCLASSOPTIONcompsoc
  \section*{Acknowledgments}
\else
  \section*{Acknowledgment}
\fi
This work was supported by the Austrian Science Fund (FWF): P25489-N23 (``KAVA-Time'') and by the NFB -- Lower Austrian Research and Education Company and the Provincial Government of Lower Austria, Dep. of Science and Research (``IntelliGait'' LSC14-005).
Cordial thanks to Marianne Worisch, Christina Niederer\new{,} and Niklas Th\"ur for their support and to Tarique Siragy for  proofreading. Special thanks to all focus group members and test participants who have agreed to volunteer in this project.

\ifCLASSOPTIONcaptionsoff
  \newpage
\fi



\bibliographystyle{IEEEtran}
\bibliography{Biblios/VisBiblio,Biblios/GaitBiblio}
%\begin{thebibliography}{1}
%\end{thebibliography}


%\end{IEEEbiography} 

\begin{IEEEbiography}[{\includegraphics[width=1in,height=1.25in,clip,keepaspectratio]{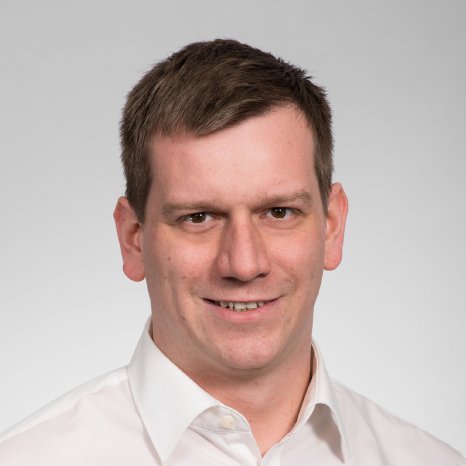}}]{Markus Wagner}
%\begin{IEEEbiographynophoto}{Markus Wagner}
%is a research associate at the Institute of Creative$\backslash{}$Media/Technologies, St. P\"olten University of Applied Sciences (Austria) and a PhD student at the Faculty of Informatics at Vienna University of Technology (Austria).
%His research activities involve knowledge-assisted visual analytics methods for behavior-based malware analysis as well as clinical gait analysis.
%Markus studied Game Engineering and Simulation at UAS Technikum Wien where he received his MSc degree in 2013 and Industrial Simulation at St. Poelten UAS where he received his BSc degree in 2011.
%is a research associate at the Institute of Creative$\backslash{}$Media/Technologies, St. P\"olten University of Applied Sciences (Austria). He holds a master degree in Game Engineering and Simulation since 2013 and is a PhD student at the Faculty of Informatics at TU Wien (Austria). His research interests involve knowledge-assisted information visualization and visual analytics methods.
is a research associate at the Institute of Creative$\backslash{}$Media/Technologies, St. P\"olten University of Applied Sciences (Austria) since 2014. He received his PhD in Computer Science from TU Wien (Austria) in June 2017 with highest distinction. His research interests involve knowledge-assisted information visualization and visual analytics methods.
\end{IEEEbiography}

\begin{IEEEbiography}[{\includegraphics[width=1in,height=1.25in,clip,keepaspectratio]{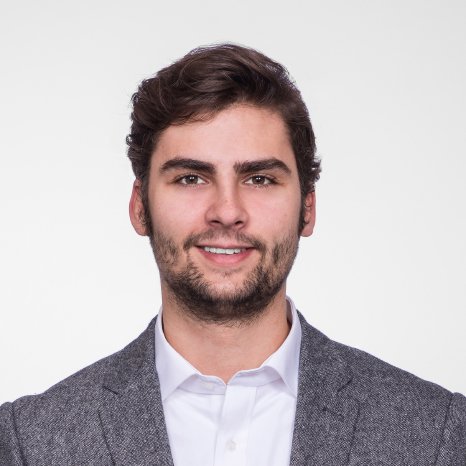}}]{Djordje Slijep\v{c}evi\'{c}}
%\begin{IEEEbiographynophoto}{Djordje Slijep\v{c}evi\'{c}}
is a research associate at the Institute of Creative$\backslash{}$Media/Technologies, St.~P\"olten University of Applied Sciences (Austria) and a PhD student at the Faculty of Informatics at TU Wien (Austria). His research focuses on feature learning, pattern recognition and machine learning. He graduated from TU Wien (Austria) with a MSc in Computer Engineering.
\end{IEEEbiography}

\begin{IEEEbiography}[{\includegraphics[width=1in,height=1.25in,clip,keepaspectratio]{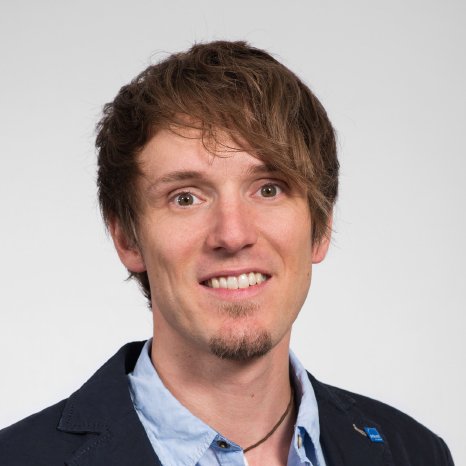}}]{Brian Horsak}
%\begin{IEEEbiographynophoto}{Brian Horsak}
holds a doctorate degree in Biomechanics and Kinesiology. He is researcher at the St. P\"olten University of Applied Sciences (Austria) in the Department of Health Sciences. His research focuses on musculoskeletal loading of the lower extremities during locomotion and on innovative technologies supporting patients and therapists during rehabilitation.
\end{IEEEbiography}

\begin{IEEEbiography}[{\includegraphics[width=1in,height=1.25in,clip,keepaspectratio]{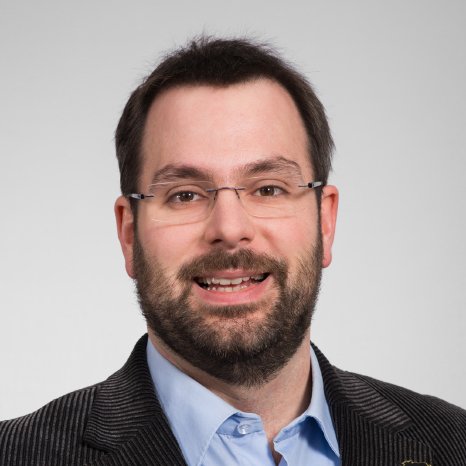}}]{Alexander Rind}
%\begin{IEEEbiographynophoto}{Alexander Rind}
is a research associate at the Institute of Creative$\backslash{}$Media/Technologies, St. P\"olten University of Applied Sciences (Austria) and a PhD student at the Faculty of Informatics at TU Wien (Austria).
Alexander studied Business Informatics in Vienna and Lund and received his MSc degree from TU Wien. % in 2004.
His research focuses on interaction and knowledge in visual analytics and visualization for health care. % of electronic health records.
% knowledge-assisted visual analytics methods,
% He is an active member of the research community and is in the program committees for EuroVA and VAHC.
\end{IEEEbiography}

\begin{IEEEbiography}[{\includegraphics[width=1in,height=1.25in,clip,keepaspectratio]{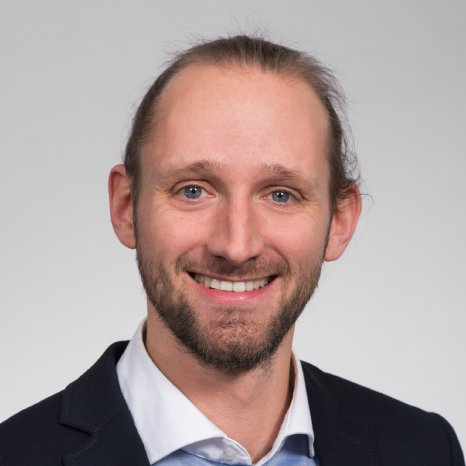}}]{Matthias Zeppelzauer}
%\begin{IEEEbiographynophoto}{Matthias Zeppelzauer}
is a senior researcher at the Institute of Creative$\backslash{}$Media/Technologies at St. P\"olten University of Applied Sciences (Austria) since 2013. He received his PhD in Computer Science from TU Wien (Austria) in 2011 with highest distinction. His research focuses on multimedia signal processing, pattern recognition and machine learning. 
\end{IEEEbiography}

\begin{IEEEbiography}[{\includegraphics[width=1.0in,height=1.25in,clip,keepaspectratio]{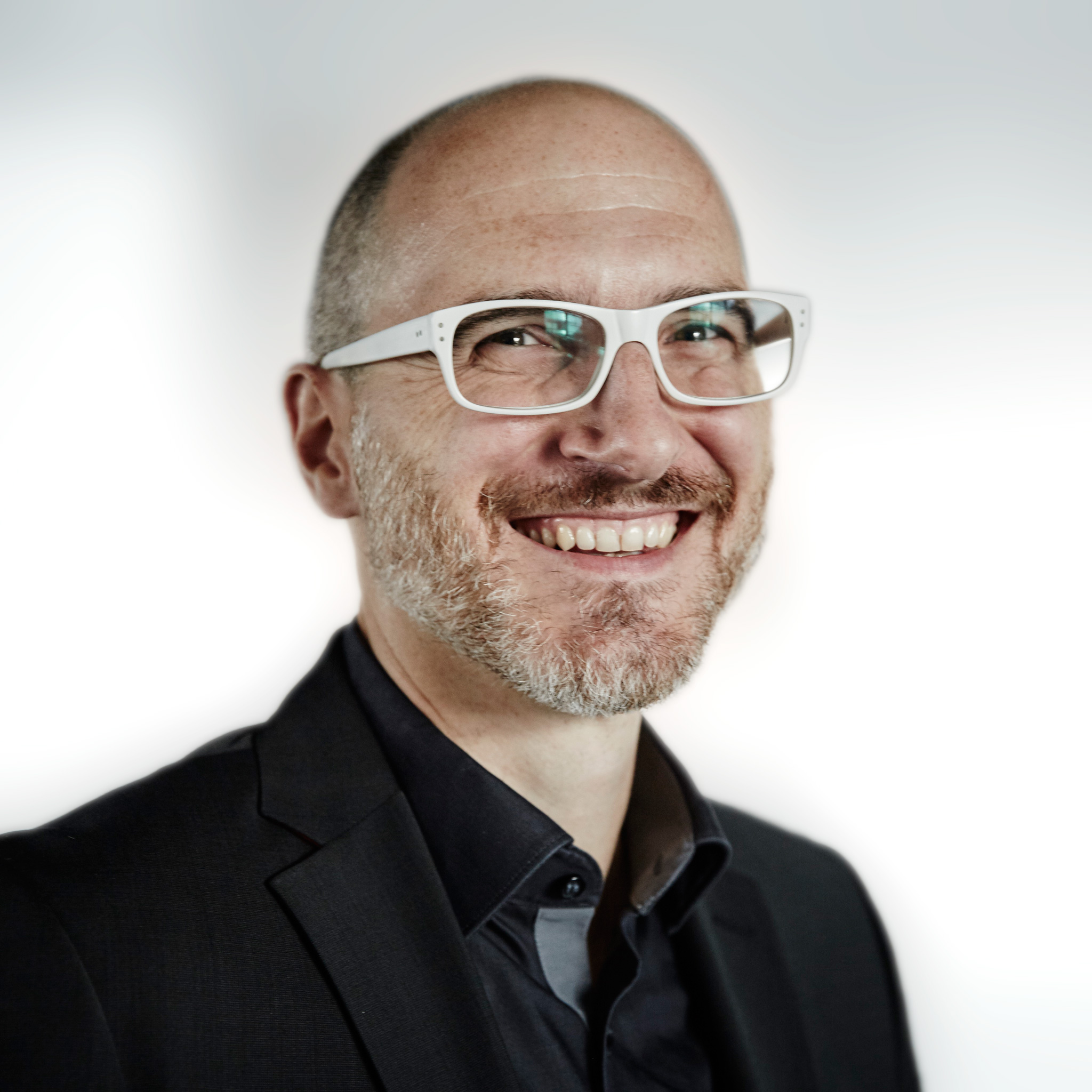}}]{Wolfgang Aigner}
%\begin{IEEEbiographynophoto}{Wolfgang Aigner}
is %a professor and 
the scientific director of the Institute of Creative$\backslash{}$Media/Technologies at St. P\"olten University of Applied Sciences (Austria) and adjunct professor at TU Wien (Austria). 
He received his PhD degree and habilitation in computer science from TU Wien in 2006 and 2013. 
His main research interests include visual analytics and information visualization with a particular focus on time-oriented data. 
%Wolfgang Aigner received his PhD degree and habilitation from TU Wien (Austria) in 2006 and 2013. 
%He authored and co-authored more than 100 peer-reviewed publications as well as the book ``Visualization of Time-Oriented Data'' (Springer, 2011) that is devoted to a systematic view on this topic.
\end{IEEEbiography}
\end{document}